\documentclass[conference]{IEEEtran}
\IEEEoverridecommandlockouts

\usepackage{cite}

\PassOptionsToPackage{hyphens}{url}
\usepackage[hidelinks]{hyperref}
\usepackage{amsmath,amssymb,amsfonts}
\usepackage{graphicx}
\usepackage{textcomp}
\usepackage{xcolor}
\usepackage{todonotes}
\usepackage{listings}
\usepackage{float}
\usepackage{enumitem}
\usepackage{wrapfig, framed}
\usepackage[export]{adjustbox}
\usepackage{physics}
\usepackage{tcolorbox}
\usepackage{comment}
\usepackage{balance}

\usepackage[linesnumbered,ruled,vlined]{algorithm2e}

\usepackage{algorithmic}

\usepackage{stmaryrd}
\usepackage{array}
\newcolumntype{H}{>{\iffalse}c<{\fi}@{}}

\usepackage[skip=5pt]{caption} % example skip set to 2pt
\usepackage{subcaption}
\definecolor{csgreen}{rgb}{0.0, 0.5, 0.0}
\definecolor{guppiegreen}{rgb}{0.0, 1.0, 0.5}
\definecolor{asparagus}{rgb}{0.27, 0.35, 0.27}
\definecolor{crimson}{HTML}{C90016}
\definecolor{tmagenta}{HTML}{FF00FF}
\definecolor{tred}{HTML}{FF0000}
\definecolor{posex}{HTML}{0ca765}
\definecolor{bottlegreen}{rgb}{0.0,0.42,0.31}

\definecolor{fuchsiapink}{rgb}{1.0, 0.47, 1.0}
\definecolor{fluorescentpink}{rgb}{1.0, 0.08, 0.58}
\definecolor{darkpink}{rgb}{0.91, 0.33, 0.5}
\definecolor{armygreen}{rgb}{0.1,0.3,0}
\definecolor{darkgreen}{rgb}{0.0, 0.3, 0.15}
\definecolor{dgreen}{rgb}{0.12, 0.3, 0.17}
\definecolor{mygreen}{rgb}{0,0.6,0}
\definecolor{mygray}{rgb}{0.5,0.5,0.5}
\definecolor{mymauve}{rgb}{0.58,0,0.82}
\definecolor{darkred}{rgb}{0.55,0,0}
\definecolor{banana}{rgb}{0.98,0.91,0.71}
\definecolor{coolblack}{rgb}{0,0.18,0.39}
\definecolor{cream}{rgb}{1,0.99,0.82}
\definecolor{emerald}{rgb}{0.31, 0.78, 0.47}

\definecolor{codegreen}{rgb}{0,0.6,0}
\definecolor{codegray}{rgb}{0.5,0.5,0.5}
\definecolor{codepurple}{rgb}{0.58,0,0.82}
\definecolor{backcolour}{rgb}{0.95,0.95,0.92}
\definecolor{aliceblue}{rgb}{0.94,0.97,0.80}
\definecolor{darkmagenta}{rgb}{0.55, 0.0, 0.55}
\definecolor{aqua}{rgb}{0.0, 1.0, 1.0}
\definecolor{blizzardblue}{rgb}{0.67, 0.9, 0.93}

\newcommand{\B}[1]{\langle #1\rangle}

\newcommand{\highlight}[1]{\colorbox{yellow}{\ensuremath{#1}}}

\newcommand{\tool}{\textsc{P\=a\d{n}ini}} %googled:  grammarian for sanskrit % please change as per your choice. 

\newcommand{\fex}[3]{\texttt{\fbox{{$\llbracket$#1$\rrbracket$}$_{x = #2}$ = #3}}}
\newcommand{\fexs}[2]{\texttt{\fbox{{$\llbracket$#1$\rrbracket$}$_{x = #2}$}}}
\newcommand{\semg}[1]{{\llbracket #1 \rrbracket}^\mathcal{G}}
\newcommand{\semb}[1]{{\llbracket #1 \rrbracket}_\beta}
\newcommand{\semgb}[1]{{\llbracket #1 \rrbracket}^\mathcal{G}_\beta}
\newcommand{\semgho}[1]{{\llbracket #1 \rrbracket}_{\mathcal{G}_1^\bullet}}
\newcommand{\hole}[1]{#1^{\bullet}}
\newcommand{\hhole}[1]{\highlight{#1^{\bullet}}}
\newcommand{\ghole}{\mathcal{G}^{\bullet}}
\newcommand{\gcomplete}[1]{\mathcal{G}^{\{#1\}}}
\newcommand{\g}{\mathcal{G}}
\newcommand{\s}[1]{\langle #1 \rangle}

\newcommand{\p}[1]{\noindent\textbf{#1. }}
\newcommand{\figref}[1]{Fig.~{\ref{#1}}}

\newcommand{\forarxiv}[1]{#1}
\newcommand{\forFMCAD}[1]{#1}

\renewcommand{\forFMCAD}[1]{}

\begin{document}
\setlength{\belowdisplayskip}{5pt}
\setlength{\abovedisplayskip}{5pt} 
\setlength{\textfloatsep}{5pt}

\title{Synthesis of Semantic Actions in Attribute Grammars}
%\author{}
%\institute{}

\author{
	\IEEEauthorblockN{Pankaj Kumar Kalita}
	\IEEEauthorblockA{\textit{Computer Science and Engineering} \\
		\textit{Indian Institute of Technology Kanpur}\\
		Kanpur, India \\
		{\small \texttt{pkalita@cse.iitk.ac.in}}}
	\and
	\IEEEauthorblockN{Miriyala Jeevan Kumar}
	\IEEEauthorblockA{
		\textit{Fortanix Tech. India Pvt. Ltd.}\\
		Bengaluru, India \\
		{\small \texttt{g1.miriyala@gmail.com}}}
	\and 
	\IEEEauthorblockN{Subhajit Roy}
	\IEEEauthorblockA{\textit{Computer Science and Engineering} \\
		\textit{Indian Institute of Technology Kanpur}\\
		Kanpur, India \\
		{\small \texttt{subhajit@iitk.ac.in}}}
	
}

\maketitle

\begin{abstract}
\textit{Attribute grammars} allow the association of \textit{semantic actions} to the production rules in context-free grammars, providing a simple yet effective formalism to define the semantics of a language. However, drafting the semantic actions can be tricky and a large drain on developer time. In this work, we propose a synthesis methodology to automatically infer the semantic actions from a set of examples associating strings to their \textit{meanings}. We also propose a new coverage metric, \textit{derivation coverage}. We use it to build a sampler to effectively and automatically draw strings to drive the synthesis engine. We build our ideas into our tool, {\tool}, and empirically evaluate it on twelve benchmarks, including a forward differentiation engine, an interpreter over a subset of Java bytecode, and a mini-compiler for C language to two-address code. Our results show that {\tool} scales well with the number of actions to be synthesized and the size of the context-free grammar, significantly outperforming simple baselines.
\end{abstract}


\section{Introduction}

Attribute grammars~\cite{Knuth1968} provide an effective formalism to supplement a language syntax (in the form of a context-free grammar) with semantic information. The semantics of the language is described using \textit{semantic actions} associated with the grammar productions. The semantic actions are defined in terms of \textit{semantic attributes} associated with the non-terminal symbols in the grammar. 

Almost no modern applications use hand-written parsers anymore; instead, most language interpretation engines today use automatic parser generators (like \textsc{yacc}~\cite{yacc}, \textsc{Bison}~\cite{bison}, \textsc{\textsc{antlr}}~\cite{antlr} etc.). These parser generators employ the simple, yet powerful formalism of attribute grammars to couple parsing with semantic analysis to build an efficient frontend for language understanding. This mechanism drives many applications like model checkers (eg. \textsc{spin}~\cite{spinRepo}), automatic theorem provers (eg. \textsc{q3b}~\cite{q3b}, \textsc{cvc5}~\cite{cvc5}),  compilers (eg. \textsc{cil}~\cite{cil}), database engines (eg. \textsc{mySQL}~\cite{mysql}) etc.%, , interpreters (eg. ???) and compilers (eg. ???). %This simple yet powerful formalism has been the {\it de facto} choice for building most of the language parsers, interpreters, and compilers for the last few decades. In fact \pankaj{i.e., parsing smt-lib~\cite{smtlibRepo}, spin model checker~\cite{spinRepo}}.

However, defining appropriate semantic actions is often not easy: they are tricky to express in terms of the inherited and synthesized attributes over the grammar symbols in the respective productions. Drafting these actions for large grammars requires a significant investment of developer time.

In this work, we propose an algorithm to automatically synthesize semantic actions from \textit{sketches} of attribute grammars. \figref{lst:motivationGrammar} shows a sketch of an attribute grammar for automatic forward differentiation using dual numbers (we explain the notion of dual numbers and the example in detail in \S\ref{sec:moti}). The production rules are shown in \textcolor{darkgreen}{green} color while the semantic actions are shown in the \textcolor{blue}{blue} color. 
Our synthesizer attempts to infer the definitions of the \textit{holes} in this sketch (the function calls $\hole{h_1}$, $\hole{h_2}$, $\hole{h_3}$, $\hole{h_4}$, $\hole{h_5}$, $\hole{h_6}$); we show these holes in \colorbox{yellow}{yellow} background. As an attribute grammar attempts to assign ``meanings" to language strings, the meaning of a string in this language is captured by the \colorbox{pink}{output} construct.

 %(shown in \colorbox{pink}{pink} color background). %\todo{find good colors}

%In this work, we propose an algorithm to automatically synthesize semantic actions from attribute grammar sketches. \figref{lst:motivationGrammar} shows a sketch of an attribute grammar for automatic forward differentiation using dual numbers (the example is explain in detail in \S\ref{sec:moti}). The production rules are shown in \textcolor{darkgreen}{green} color while the semantic actions are shown in the \textcolor{blue}{blue} color. The \textit{holes} in this sketch that need to be synthesized are shown as function calls marked with an \colorbox{yellow}{yellow} background. As an attribute grammar attempts to assign ``meanings" to language strings, the meaning of a string in this language is captured by the \colorbox{pink}{output} construct (shown in \colorbox{pink}{pink} color background). %\todo{find good colors}

\begin{figure}[t]

%\begin{minipage}{\textwidth}
%	\vspace{1em}
%\begin{minipage}{.5\textwidth}

	\small

	\begin{tabular}{>{\color{darkgreen}}l >{\color{blue}}l}
		\textsf{S} $\rightarrowtail$  \textsf{E} $^{[1]}$ &\hspace*{2mm}	{\colorbox{pink}{output(\textsf{E}.val;)}}\\
		\textsf{E} $\rightarrowtail$ \textsf{E} + \textsf{F} $^{[2]}$	&\hspace*{2mm}    {\textsf{E}.val $\leftarrow$ \colorbox{yellow}{$\hole{h_1}$(\textsf{E}.val, \textsf{F}.val);}}\\
		\hspace*{6mm}\textbar\  \textsf{E} - \textsf{F} $^{[3]}$ &\hspace*{2mm} {\textsf{E}.val $\leftarrow$ \colorbox{yellow}{$\hole{h_2}$(\textsf{E}.val, \textsf{F}.val);}}\\
		\hspace*{6mm}\textbar\  \textsf{F} $^{[4]}$	&\hspace*{2mm}	{\textsf{E}.val $\leftarrow$ \textsf{F}.val;}\\
		\textsf{F} $\rightarrowtail$ \textsf{F} * \textsf{K} $^{[5]}$&\hspace*{2mm}  {\textsf{F}.val $\leftarrow$ \colorbox{yellow}{$\hole{h_3}$(\textsf{F}.val, \textsf{K}.val);}}\\
		\hspace*{6mm}\textbar\  \textsf{K}  	$^{[6]}$  &\hspace*{2mm}     {\textsf{F}.val $\leftarrow$ \textsf{K}.val;}\\
		\textsf{K} $\rightarrowtail$ \textsf{K} \textasciicircum \textsf{num} $^{[7]}$ &\hspace*{2mm} {\textsf{K}.val $\leftarrow$ \colorbox{yellow}{$\hole{h_4}$(\textsf{K}.val, \textsf{num});}}\\
		\hspace*{6mm}\textbar\  \textsf{SIN ( K )} $^{[8]}$ &\hspace*{2mm}   {\textsf{K}.val $\leftarrow$ \colorbox{yellow}{$\hole{h_5}$(\textsf{K}.val);}}\\
		\hspace*{6mm}\textbar\  \textsf{COS ( K )}  $^{[9]}$&\hspace*{2mm} {\textsf{K}.val $\leftarrow$ \colorbox{yellow}{$\hole{h_6}$(\textsf{K}.val);}}\\
		\hspace*{6mm}\textbar\ \textsf{num} $^{[10]}$&\hspace*{2mm}	{\textsf{K}.val $\leftarrow$ getVal(\textsf{num}) + 0$\varepsilon$;}\\
		\hspace*{6mm}\textbar\  \textsf{var} $^{[11]}$	&\hspace*{2mm} 	{\textsf{K}.val $\leftarrow$ lookUp($\Omega$, \textsf{var}) + 1$\varepsilon$;} 
	\end{tabular} 
	\captionof{figure}{\footnotesize Attribute grammar for automatic forward differentiation \\($\Omega$ is the symbol table)  \label{lst:motivationGrammar}}

%\end{minipage}
%\hfill
%\begin{minipage}{.45\textwidth}
%	\centering
%		\includegraphics[scale=.48]{fig/parseTree.eps}
%		\captionof{figure}{Parse tree for input \texttt{x\textasciicircum 2+4*x+5}\label{fig:parseTree}}	
%\end{minipage}
%\vspace{1em}
%\end{minipage}
\end{figure}

This is a novel synthesis task: the current program synthesis tools synthesize a program such that a desired specification is met. In our present problem, we attempt to synthesize semantic actions within an attribute grammar: the synthesizer is required to \textit{infer} definitions of the holes such that for \textit{all} strings in the language described by the grammar, the computed semantic value (captured by the \colorbox{pink}{output} construct) matches the intended semantics of the respective string---this is a new problem that cannot be trivially mapped to a program synthesis task. 

\begin{figure}[t]%{l}{.23\textwidth}
	\centering
	\includegraphics[scale=.55]{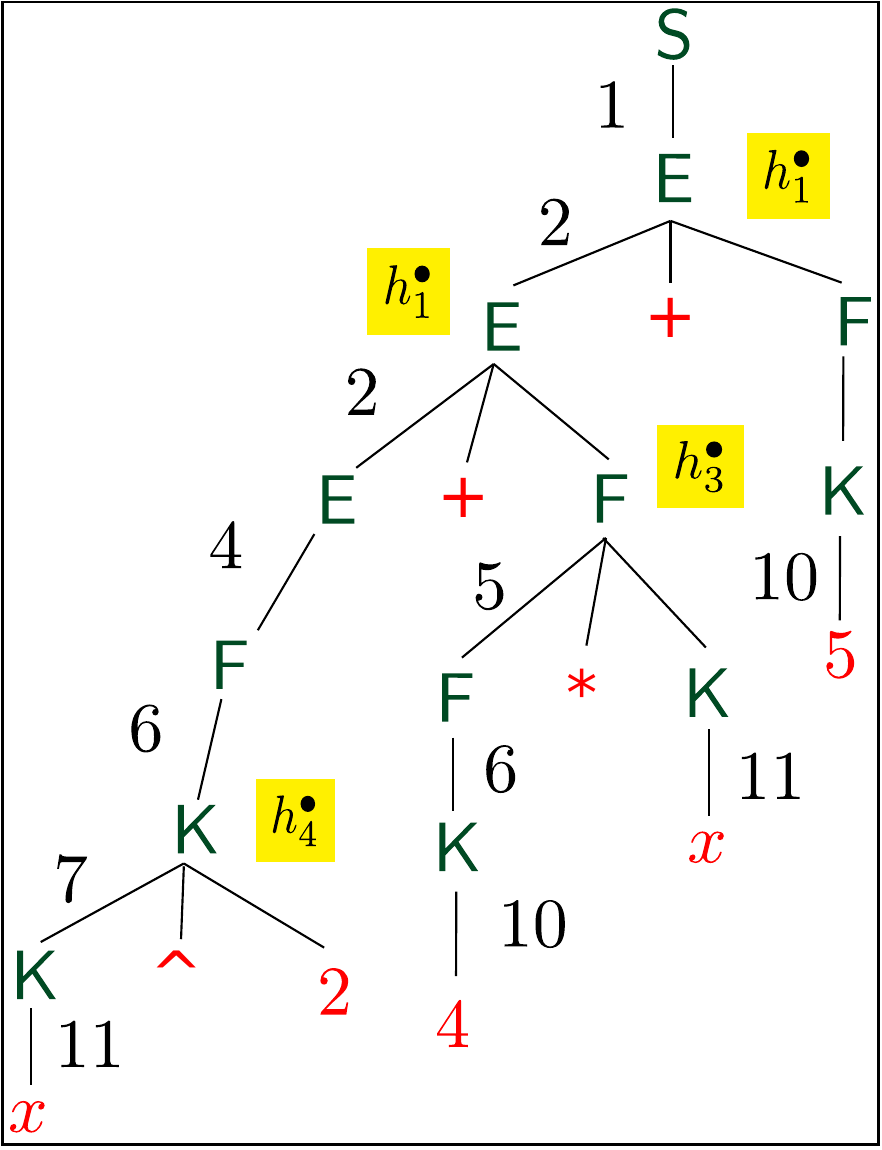}
	\captionof{figure}{Parse tree for input \texttt{x\textasciicircum 2+4*x+5}\label{fig:parseTree}}	
\end{figure}

Our core observation to solve this problem is the follows: for any string in the language, the sequence of semantic actions executed for the syntax-directed evaluation of any string is a loop-free program. This observation allows us to reduce attribute grammar synthesis \textit{to a set program synthesis tasks}. Unlike a regular program synthesis task where we are interested in synthesizing a single program, the above reduction requires us to solve a set of \textit{dependent} program synthesis instances simultaneously. These program synthesis tasks are dependent as they contain common components (to be synthesized) shared by multiple programs, and hence, they cannot be solved in isolation---as the synthesis solution from one instance can influence others. 

Given a set of examples $E$ (strings that can be derived in the provided grammar) and their expected semantics $O$ (semantic values in \colorbox{pink}{output}); for each example $e_i \in E$ , we apply the reduction by sequencing the set of semantic actions for the productions that occur in the derivation of $e_i$. This sequence of actions forms a loop-free program $P_i$, with the expected semantic output $o_i \in O$ as the specification. We collect such programs-specification pairs, $\langle P_i, o_i\rangle$, to create a set of dependent synthesis tasks. An attempt at \textit{simultaneous synthesis} of all these set of tasks by simply conjoining the synthesis constraints does not scale.%constructs a symbolic trace of the computation of the semantic value corresponding to each example and constrains the output to the expected value (provided in the example). 

%Then, it attempts to synthesize definitions for all holes such that all constraints are satisfied. However, such an algorithm does not scale well. 
Our algorithm adopts an incremental, counterexample guided inductive synthesis (CEGIS) strategy, that attempts to handle only a ``small" set of programs simultaneously---those that violate the current set of examples. Starting with only a single example, the set of satisfied examples are expanded incrementally till the specifications are satisfied over all the programs in the set.

Furthermore, to relieve the developer from providing examples, we also propose an example generation strategy for attribute-grammars based on a new coverage metric. Our coverage metric, \textit{derivation coverage}, attempts to capture distinct behaviors due to the presence or absence of each of the semantic actions corresponding to the syntax-directed evaluation of different strings. 
%: we randomly select \textit{tests} from the provided set of examples. first attempt at the synthesis algorithm constructs symbolic traces and attempts to synthesize definitions for its holes such that it satisfies each of the concrete traces. A verification check attempts to the final verification discharges definitions for the required functions (Figure ???).

\begin{figure*}[t]
	\centering
	\begin{subfigure}{.3\textwidth}\centering
		\DontPrintSemicolon
		\SetAlgoNoLine
		\SetKwProg{add}{$\hole{h_1}$}{:}{}
		\add{($a_1 + a_2 \varepsilon$, $b_1 + b_2 \varepsilon$)}{
			$r \gets a_1 + b_1$\;
			
			$d \gets a_2 + b_2$ \;
			
			\KwRet $r + d\varepsilon$	
		}
		\caption{ \label{fig:addDef}}
	\end{subfigure}%
	\hfill
	\begin{subfigure}{.3\textwidth}\centering
		\DontPrintSemicolon
		\SetAlgoNoLine
		\SetKwProg{sub}{$\hole{h_2}$}{:}{}
		\sub{($a_1 + a_2 \varepsilon$, $b_1 + b_2 \varepsilon$)}{
			$r \gets a_1 - b_1$\;
			
			$d \gets a_2 - b_2$ \;
			
			\KwRet $r + d\varepsilon$	
		}
		\caption{\label{fig:subDef}}
	\end{subfigure}%
	\hfill
	\begin{subfigure}{.35\linewidth}\centering
		\DontPrintSemicolon
		\SetAlgoNoLine
		\SetKwProg{mul}{$\hole{h_3}$}{:}{}
		\mul{($a_1 + a_2 \varepsilon$, $b_1 + b_2 \varepsilon$)}{
			$r \gets a_1 * b_1$\;
			
			$d \gets a_2 * b_1 + a_1 * b_2$ \;
			
			\KwRet $r + d\varepsilon$	
		}
		%			\end{subfigure}
		\caption{ \label{fig:mulDef}}
	\end{subfigure}
	
	%	\caption{Synthesized definitions for $\hole{h_1}, \hole{h_2}, \hole{h_3}$ \label{fig:addDef}}
	%\end{figure*}
	
	%	\begin{figure}[t]
	%\newline
	%	\hspace*{2.5cm}
	\begin{subfigure}{.25\textwidth}\centering
		\DontPrintSemicolon
		\SetAlgoNoLine
		\SetKwProg{add}{$\hole{h_5}$}{:}{}
		\add{($a_1 + a_2 \varepsilon$)}{
			$r \gets sin(a_1)$\;
			
			$d \gets a_2 * cos(a_1)$\;
			
			\KwRet $r + d\varepsilon$	
		}
		\caption{\label{fig:sinDef}}
	\end{subfigure}%
	\hfill
	%	\hspace{1.5cm}
	\begin{subfigure}{.3\textwidth}\centering
		\DontPrintSemicolon
		\SetAlgoNoLine
		\SetKwProg{sub}{$\hole{h_6}$}{:}{}
		\sub{($a_1 + a_2 \varepsilon$)}{
			$r \gets cos(a_1)$\;
			
			$d \gets a_2 * sin(a_1) * -1$\;
			
			\KwRet $r + d\varepsilon$	
		}
		\caption{\label{fig:cosDef}}
	\end{subfigure}%
	\hfill
	\begin{subfigure}{.35\textwidth}\centering
		\DontPrintSemicolon
		\SetAlgoNoLine
		\SetKwProg{pow}{$\hole{h_4}$}{:}{}
		\pow{($a_1 + a_2\varepsilon$, $c$)}{
			$r \gets pow(a_1,\ c)$\;
			
			$d \gets a_2 * pow(a_1, c - 1)$ \;
			
			\KwRet $r + d\varepsilon$	
		}
		\caption{\label{fig:powDef}}
	\end{subfigure}
	\caption{Synthesized holes for holes in \figref{lst:motivationGrammar}\label{fig:correctSol}}
\end{figure*}

We build  an implementation, {\tool}\footnote{{\tool} (\includegraphics[scale=.12]{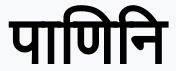}) was a Sanskrit grammarian and scholar in ancient India.}, that is capable of automatically synthesizing semantic actions (across both synthesized and inherited attributes) in attribute grammars. For the attribute grammar sketch in \figref{lst:motivationGrammar}, {\tool} automatically synthesizes the definitions of holes as shown in~\figref{fig:correctSol} in a mere 39.2 seconds. We evaluate our algorithm on a set of attribute grammars, including a Java bytecode interpreter and a mini-compiler frontend. Our synthesizer takes a few seconds on these examples.

%While there has been work on automatically synthesizing grammars and parsers~\cite{cyclops}, 
To the best of our knowledge, ours is the first work at automatic synthesis of semantic actions on attribute grammars. The following are our contributions in this work:

\begin{itemize}[noitemsep,nolistsep]
	\item We propose a new algorithm for synthesizing semantic actions in attribute grammars;
	\item We define a new coverage metric, \textit{derivation coverage}, to generate effective examples for this synthesis task;
	\item We build our algorithms into an implementation, {\tool}, to synthesize semantic actions for attribute grammars;
	\item We evaluate {\tool} on a set of attribute grammars to demonstrate the efficacy of our algorithm. We also undertake a case-study on the attribute grammar of the parser of the \textsc{spin} model-checker to automatically infer the constant-folding optimization and abstract syntax tree construction.
\end{itemize}

    \forFMCAD{An extended version of this article is available~\cite{archivePanini}.}
    The implementation and benchmarks of {\tool} are available at \href{https://github.com/pkalita595/Panini}{\it https://github.com/pkalita595/Panini}.

\section{Preliminaries}
Attribute grammars~\cite{Knuth1968} provide a formal mechanism to capture language semantics by extending a context-free grammar with \textit{attributes}. An attribute grammar $\mathcal{G}$ is specified by $\langle \mathsf{S}, P, T, N, F, \Gamma\rangle$, where %\todo{fix the fonts throughout}
%\todo{space is too less i guess}
\begin{itemize}[noitemsep,nolistsep]
    \item $T$ and $N$ are the set of \textit{terminal} and \textit{non-terminal} symbols (resp.), and $\mathsf{S}\in N$ is the \textit{start symbol};
    \item A set of (context-free) productions, $p_i \in P$, where $p_i: \mathsf{X_i} \rightarrowtail \mathsf{Y_{i1}}\mathsf{Y_{i2}}\dots\mathsf{Y_{in}}$; a production consists of a \textit{head} $\mathsf{X_i} \in N$ and \textit{body} $\mathsf{Y_{i1}}\dots\mathsf{Y_{in}}$, such that each $Y_{ik} \in T \cup N$.
    \item A set of \textit{semantic actions} $f_i \in F$;
    \item $\Gamma: P\rightarrow F$ is a map from the set of productions $P$ to the set of \textit{semantic actions} $f_i \in F$. 
\end{itemize}

%\begin{wrapfigure}{r}{.35\textwidth}
%	\includegraphics[scale=.5]{fig/parseTree.eps}
%	\caption{Parse tree for the input $x^2 + 4x+5$\label{fig:parseTree}}
%	\vspace*{-2em}
%\end{wrapfigure}

The set of productions in $\mathcal{G}$ describes a \textit{language} (denoted as $\mathcal{L(G)}$) to capture the set of \textit{strings} that can be \textit{derived} from $\textsf{S}$. A \textit{derivation} is a sequence of applications of productions $p_i \in P$ that transforms \textsf{S} to a string, $w \in \mathcal{L(G)}$; unless specified, we will refer to the \textit{leftmost} derivation where we always select the leftmost non-terminal for expansion in a sentential form. As we are only concerned with parseable grammars, we constrain our discussion in this paper to \textit{unambiguous} grammars.

The semantic actions associated with the grammar productions are defined in terms of \textit{semantic attributes} attached to the non-terminal symbols in the grammar. Attributes can be \textit{synthesized} or \textit{inherited}: while a synthesized attributes are computed from the children of a node in a parse tree, an inherited attribute is defined by the attributes of the parents or siblings. 

\figref{lst:motivationGrammar} shows an attribute grammar with \textcolor{darkgreen}{context-free productions} and the associated \textcolor{blue}{semantic actions}. \figref{fig:parseTree} shows the parse tree of the string $x^2 + 4x + 5$ on the provided grammar; each internal node of the parse tree have associated semantic actions (we have only shown the ``unknown" actions that need to be inferred). 

%Hoare triple~\cite{hoare}, \textit{\{P\} S \{Q\}} states that, if pre-condition \textit{P} is satisfied before execution of statement \textit{S}, then after executing \textit{S} post-condition \textit{Q} will also be satisfied.

\textit{Parser generators}~\cite{yacc} accept an attribute grammar and automatically generate parsers that perform a \textit{syntax-directed evaluation} of the semantic actions. For ease of discussion, we assume that the semantic actions are \textit{pure} (i.e. do not cause side-effects like printing values or modifying global variables) and generate a deterministic \textit{output value} as a consequence of applying the actions.% We denote the output \textit{value} generated by the syntax-directed \textit{semantic evaluation} of a string $w\in\mathcal{L(G)}$ as $\semg{w}$. \todo{First time defined $\semg{w}$}

An attribute grammar is \textit{non-circular} if the dependencies between the attributes in every syntax tree are acyclic. Non-circularity is a sufficient condition that all strings have unique evaluations~\cite{AGcircularTesting}. 

%    Efficient parser generators like yacc/bison~\cite{yacc} only handle L-attributed grammars. For a given production, $\mathsf{X} \rightarrowtail \mathsf{Y_1}\mathsf{Y_2}\dots\mathsf{Y_n}$, an L-attributed grammar imposes the following restrictions:
%\begin{itemize}
%    \item the synthesized attributes of $\mathsf{X}$ can only be computed from the attributes of $\mathsf{Y_1}, \dots, \mathsf{Y_n}$ ;
%    \item the inherited attributes of $\mathsf{Y_i}$ are computed from the inherited attributes of $\mathsf{X}$ and the attributes of only its \textit{left} siblings, $\mathsf{Y_1}, \dots, \mathsf{Y_{i-1}}$.
%\end{itemize}
\p{Notations} We notate production symbols by serif fonts, non-terminal symbols (or placeholders) by capital letters (eg. \textsf{X}) and terminal symbols by small letters (eg. \textsf{a}). Sets are denoted in capital letters. We use arrows with tails ($\rightarrowtail$) in productions and string derivations to distinguish it from function maps. We use the notation $e[g_1/g_2]$ to imply that all instances of the subexpression $g_2$ are to be substituted by $g_1$ within the expression $e$. We use the notation of Hoare logic~\cite{hoare} to capture program semantics: $\{P\} Q \{R\}$ implies that if the program $Q$ is executed with a \textit{precondition} $P$, it can only produce an output state in $R$; $P$ and $R$ are expressed in some base logic (like first-order logic).

\section{Overview}
\p{Sketch of an attribute grammar} We allow the \textit{sketch} $\ghole$ of an attribute grammar\footnote{as syntax directed definition (SDD)}, $\ghole=\langle \mathsf{S}, P, T, N, \hole{H}, \Gamma\rangle$, to contain \textit{holes} for unspecified functionality within the semantic actions $\hole{h_i} \in \hole{H}$. 
For example, in Fig.~\ref{lst:motivationGrammar}, the set of holes comprises of the functions $H=\{\hole{h_1}, \hole{h_2}, \hole{h_3}, 
%\hole{div}, 
\hole{h_4}, \hole{h_5}, \hole{h_6} \}$. 
If the semantic action corresponding to a production $p$ contains hole(s), we refer to the production $p$ as a \textit{sketchy} production; when the definitions for all the holes in a sketchy production are resolved, we say that the production is \textit{ready}. The \textit{completion} (denoted $\gcomplete{f_1, \dots, f_n}$) of a grammar sketch $\ghole$ denotes the attribute grammar where a set of functions $f_1, \dots, f_n$ replace the holes $\hole{h_1}, \dots, \hole{h_n}$. We denote the syntax-directed evaluation of a string $w$ on an attribute grammar $\g$ as $\semg{w}$; we consider that any such evaluation results in a \textit{value} (or $\bot$ if $w \notin \g$).

\p{Example Suite} An \textit{example} (or \textit{test}) for an attribute grammar $\g$ can be captured by a tuple $\langle w, v\rangle$ such that $w\in \mathcal{L(G)}$ and $\semg{w}=v$. A set of such examples constitutes an \textit{example suite} (or \textit{test suite}). %Given a set of examples, we construct an \textit{example suite} by partitioning the examples by derivation congruence.

If the language described by the grammar $\g$ supports \textit{variables}, then any evaluation of $\g$ needs a \textit{context}, $\beta$, that binds the free variables to \textit{input values}. We denote such examples as $\semgb{w} = v$. When the grammar used is clear from the context, we drop the superscript and simplify the notation to $\semb{w} = v$.
Consider the example \fex{x\textasciicircum 3}{2}{8 + 12$\varepsilon$}, where ``\texttt{x\textasciicircum 3}" is a string from the grammar shown in Fig.~\ref{lst:motivationGrammar} and the input string evaluates to \texttt{{8 + 12$\varepsilon$}} under the binding \texttt{x = 2}. Clearly, if the language does not support variables, the context $\beta$ is always empty.
%\todo{add example-> \pankaj{Done}}
%\todo{Put problem statement in a box}

\begin{tcolorbox}[colback=red!15!white,colframe=red!5!white]
    \p{Problem Statement} Given a sketch of an attribute grammar, $\hole{G}$, an example set $E$ and a domain-specific language (DSL) $D$, synthesize instantiations of the holes by strings, $w \in D$, such that the resulting attribute grammar agrees with all examples in $E$.
\end{tcolorbox}

In other words, {{\tool}} synthesizes functions $f_1, \dots, f_n$ in the domain-specific language $D$ such that the completion $\gcomplete{f_1, \dots, f_n}$ satisfies all examples in $E$. 

\subsection{Motivating example: Automated Synthesis of a Forward Differentiation Engine \label{sec:moti}}
We will use synthesis of an automatic forward differentiation engine using dual numbers~\cite{dualNumbers} as our motivating example. We start with a short tutorial on how dual numbers are used for forward differentiation.
%\todo{write how forward differentiation is done using dual numbers with an example}

\subsubsection{Forward Differentiation using Dual numbers\label{sec:dualNo}}
%Dual number is of the form $a + b\varepsilon$, where both $a$ and $b$ are real numbers along with the property $\varepsilon^2 = 0$. $a$ is called as \textit{real} part and $b$ is known as $dual$ part as $\varepsilon$ is associated with it. It is similar to the complex number of type $c + di$, where $c, d$  are real numbers and $i^2 = -1$.

Dual numbers, written as $a + b\varepsilon$, captures both the value of a function $f(x)$ (in the \textit{real} part, $a$), and that its differentiation with respect to the variable $x$, $f'(x)$, (in the \textit{dual} part,~$b$)---within the same number. Clearly, $a, b \in \mathbb{R}$ and we assume $\varepsilon^2 = 0$ (as it refers to the second-order differential, that we are not interested to track). The reader may draw parallels to complex numbers that are written as $a + ib$, where `$i$' identifies the imaginary part, and $i^2 = -1$.

Let us understand forward differentiation by calculating $f'(x)$ at $x = 3$ for the function $f(x) = x^2 + 4x + 5$. 

First, the term $x$ needs to be converted to a dual number at $x = 3$. For $x=3$, the real part is clearly $3$. To find the dual part, we differentiate the term with respect to variable $x$, i.e. {\Large $\frac {dx}{dx}$} that evaluates to $1$. Hence, the dual number representation of the term $x$ at $x = 3$ is $3 + 1\varepsilon$. 

Now, dual value of the term $x^2$ can be computed simply by taking a square of the dual representation of $x$: 
\[	\overbrace{(3 + 1\varepsilon)^2}^{x^2} = \overbrace{(3 + 1\varepsilon)}^{x} * \overbrace{(3 + 1\varepsilon)}^{x} = 3^2 + (2*3*\varepsilon) + \varepsilon^2  = 9+6\varepsilon + 0 \] 

Finally, the dual number representation for the constant $4$ is $4 + 0\varepsilon$ (as differentiation of constant is $0$). Similarly, the dual value for $4x$: $(4+0\varepsilon) * (3+1\varepsilon) = 12+4\varepsilon+0$. So, we can compute the dual number for $f(x) = x^2 + 4x + 5$ as: 
\[\overbrace{(9+6\varepsilon)}^{x^2} + \overbrace{(12+4\varepsilon)}^{4x} + \overbrace{(5 + 0\varepsilon)}^{5} = 26 + 10\varepsilon\]

Hence, the value of $f(x)$ at $x=3$ is $f(3) = 26$ (real part of the dual number above) and that of its derivative, $f'(x) = 2x+4$ is $f'(3) = 10$, which is indeed given by the the dual part for the dual number above.

\subsubsection{Synthesizing a forward differentiation engine} The attribute grammar in \figref{lst:motivationGrammar} (adapted from~\cite{forwardDiff}) implements forward differentiation for expressions in the associated context-free grammar; we will use this attribute grammar to illustrate our synthesis algorithm. lookUp($\Omega$, \textsf{var}) returns the value of the variable \textsf{var} from symbol table $\Omega$.
% This attribute grammar accepts strings the form \fexs{expr}{c}, where $c$ is the binding for variable $x$ in the input expression \texttt{expr}, i.e., \fexs{x\textasciicircum 2 + 2*x}{2}. %Attributes of non-terminals is of struct type, containing both real and dual part.  

We synthesize programs for the required functionalities for the holes from the domain-specific language (DSL) shown in Equation~\ref{Eq:DSLmotivation}. Function \texttt{pow($a, c$)} calculates  $a$ raised to the power of $c$. 
We assume the availability of an input-output oracle, $Oracle(w\B{\beta})$, that returns the expected semantic value for string $w$ under the context $\beta$.

\begin{equation}
\label{Eq:DSLmotivation}
% \begin{footnotesize}
\begin{array}{r@{\hspace{1.0ex}}c@{\hspace{1.0ex}}l}
\small
%\textit{Fun} & ::= & \lambda \texttt{$a_1$,$a_2$}\ .\ \langle E, E \rangle \  |\ \lambda \texttt{$a, num$}\ .\ \langle F, F \rangle \\
\textit{Fun} & ::= & C + C\varepsilon \  \\
C & ::= & \texttt{$var$} \mid \texttt{$num$}\ |\ 1\ |\ 0\ |\ {-}C\ 
		|\  C + C\ |\ C - C\ |\ C * C\ \\
		&& |\ \texttt{sin}(C) \mid \texttt{cos}(C) \mid \texttt{pow}(C,\ num)\ %.\texttt{real} \ |\ F.\texttt{dual}
%F & ::= & \texttt{$a$}\ |\ \texttt{$num$}\ |\ 1\ |\ 0\ |\ {-}F\ 
%|\  F + F\ |\ F - F\ |\ F * F\ \\
%		&& |\ \texttt{pow}(F,\ F)\ |\ F %.\texttt{real} \ |\ F.\texttt{dual}
%.\texttt{real} \ |\ E.\texttt{dual}\\
\end{array}
% \end{footnotesize}                
\end{equation}

%In the beginning we will generate a set of inputs for the grammar as shown in Algorithm~\ref{algo:genex}. It generates following set of inputs.
%, where the value in the form of strings $w$ and value $v$ resulting from the semantic evaluation of $w$ in the context $\beta$. 
% \fex{x \textasciicircum\ 2 + 2 * x + 5}{2}{13 + 6$\varepsilon$}: here $f(x) = x\ \hat{}\  2 + 2 * x + 5$ \todo{explain example of why one of the examples is correct; use an example that spans multiple holes}. 
%Hence, the example indeed provides insights about the semantics of forward differentiation.
% \todo{please check this example }

\subsection{Synthesis of semantic actions}
\begin{wrapfigure}{r}{0.39\linewidth}
	\vspace{-7ex}
	\begin{framed}
		\DontPrintSemicolon
		\SetAlgoNoLine
		\SetKwProg{mul}{$\hole{h_3}$}{:}{}
		\mul{($a_1 + a_2 \varepsilon$, $b_1 + b_2 \varepsilon$)}{
			$r \gets a_1 * b_1$\;
			
			%	$d \gets a_2 * b_1 + a_1$ \;
			$d \gets  b_1 + b_2 + 3 * a_2$ \; % a_1 + a_2 * b_1$ \;%2 * b_1 - 2$ \;
			
			\KwRet $r + d\varepsilon$	
		}
	\end{framed}
%	\vspace{-2ex}
	\caption{Wrong definition of $\hole{h_3}$ \label{fig:wrongMulDef}}	
%	\vspace{-5ex}
\end{wrapfigure}
Our core insight towards solving this synthesis problem is that the sequence of semantic actions corresponding to the syntax-directed evaluation of any string on the attribute grammar constitutes a loop-free program. 
%Hence, the constraint that must be satisfied for the above example can be expressed as the following Hoare triple\footnote{Hoare triple \textit{\{P\} S \{Q\}} states that, if pre-condition \textit{P} is satisfied before execution of statement \textit{S}, then after executing \textit{S} post-condition \textit{Q} will also be satisfied.}: \todo{This sentence need to change}

\figref{fig:hoareCons} shows the loop free program from the semantic evaluation of the example \fex{x\textasciicircum 2+4*x+5}{3}{26 + 10$\varepsilon$}; the Hoare triple captures the synthesis constraints over the holes.

Similarly, our algorithm constructs constraints (as Hoare triples) over the set of all examples $E$ (e.g. 
{	\ttfamily 
%	\fbox{2 : x \textasciicircum\ 3 = 8 + 12$\varepsilon$}, 
	\fex{x+x}{13}{26 + 2$\varepsilon$}, \fex{3-x}{7}{-4 - 1$\varepsilon$}, \\ \fex{x*x}{4}{16 + 8$\varepsilon$},\\
	 \fex{sin(x\textasciicircum 2)}{3}{0.41 - 5.47$\varepsilon$},\\ 
	 \fex{cos(x\textasciicircum 2)}{2}{-0.65 + 3.02$\varepsilon$}, \fex{x*cos(x)}{4}{-2.61 + 2.37$\varepsilon$}}).

Synthesizing definitions for holes that satisfy Hoare triple constraints of  all the above examples yields a valid completion of the sketch of the attribute grammar (see~\figref{fig:correctSol}). As the above queries are ``standard" program synthesis queries, they can be answered by off-the-shelf program synthesis tools \cite{sketch, rosette}. Hence, our algorithm reduces the problem of synthesizing semantic actions for attribute grammars to solving a conjunction of program synthesis problems.%\todo{Here I removed the name of sketch and rosette and just cited them, otherwise last word comes in the next line}

\begin{wrapfigure}{r}{0.53\linewidth}
	\vspace*{-2em}
	\hspace{-1cm}
	\begin{framed}	
		\hspace*{-.8em}
		\begin{minipage}{0.39\linewidth}

		\colorbox{blizzardblue}{$\mathtt{\{x = 3\}}$}\\
		$\mathtt{K_1.val \gets 3 + 1\varepsilon;}$\\
		$\mathtt{K_2.val \gets \hole{h_4}(K_1.val, 2);}$\\
		$\mathtt{K_3.val \gets 4 + 0\varepsilon;}$\\
		$\mathtt{F_1.val \gets \hole{h_3}(K_3.val, K_1.val);}$\\
		$\mathtt{E_1.val \gets \hole{h_1}(K_2.val, F_1.val);}$\\
		$\mathtt{K_4.val\gets 5 + 0\varepsilon;}$\\
		$\mathtt{output \gets \hole{h_1}(E_1.val, K_4.val);}$\\
		\colorbox{blizzardblue}{$\mathtt{\{output = 26+10\varepsilon\}}$} 
		\end{minipage}
	\end{framed}
	\caption{Hoare triple constraint for $x^2+4x+5$ at $x=3$\label{fig:hoareCons}}
	\vspace*{-1em}
\end{wrapfigure}

While the above conjunction can be easily folded into a single program synthesis query and offloaded to a program synthesis tool, quite understandably, it will not scale. To scale the above problem, we employ a refutation-guided inductive synthesis procedure: we sort the set of examples by increasing complexity, completing the holes for the easier instances first. The synthesized definitions are \textit{frozen} while handling new examples; however, unsatisfiability of a synthesis call with frozen procedures \textit{refutes} the prior synthesized definitions. Say  we need to synthesize definitions for $\{\hole{h_0}, \dots, \hole{h_9}\}$ and examples $\{e_1, \dots, e_{i-1}\}$ have already been handled, with definitions $\{\hole{h_1}=f_1, \dots, \hole{h_5}=f_5\}$ already synthesized. To handle a new example, $e_i$, we issue a synthesis call for procedures $\{\hole{h_6}, \dots, \hole{h_9}\}$ with definitions $\{\hole{h_1}=f_1, \dots, \hole{h_5}=f_5\}$ frozen. Say, the constraint corresponding to $e_i$ only includes calls $\{\hole{h_2}, \hole{h_4}, \hole{h_6}, \hole{h_8}\}$ and the synthesis query is unsatisfiable. In this case, we unfreeze \textit{only} the participating frozen definitions (i.e. $\{\hole{h_2}, \hole{h_4}\}$) and make a new synthesis query. As new query only attempts to synthesize a few new calls (with many participating calls frozen to previously synthesized definitions), this algorithm scales well.

For example, consider the grammar in~\figref{lst:motivationGrammar}: the loop-free program resulting from the semantic evaluation of the input \fex{x+x}{13}{26 + 2$\varepsilon$} (say trace $t_1$) includes only one $\hole{h_1}$. Hence, we synthesize $\hole{h_1}$ with only the constraint $\{x=13\} t_1 \{output=26 + 2\varepsilon\}$, that results in the definition shown in \figref{fig:addDef}. Next, we consider the input \fex{x*x}{4}{16 + 8$\varepsilon$}; its constraint includes the holes $\hole{h_3}$, which is synthesized as the function definition shown in~\figref{fig:wrongMulDef}. Now, with $\{\hole{h_1}, \hole{h_3}\}$ frozen to their respective synthesized definitions, we attempt to handle \fex{x*cos(x)}{4}{-2.61 + 2.37$\varepsilon$}. Its constraint includes the holes \{$\hole{h_3}$, $\hole{h_6}$\}; now we \textit{only} attempt to synthesize $\hole{h_6}$ while constraining $\hole{h_3}$ to use the definition in~\figref{fig:wrongMulDef}. In this case the synthesizer fails to synthesize $\hole{h_6}$ since the synthesized definition for $\hole{h_3}$ is incorrect, thereby \textit{refuting} the synthesized definition of $\hole{h_3}$. Hence, we now unfreeze $\hole{h_3}$ and call the synthesis engine again to synthesize both $\hole{h_3}$ and $\hole{h_6}$ together. This time we succeed in inferring correct synthesized definition as shown in~\figref{fig:correctSol}.

\subsection{Example Generation}

%We also propose a new coverage metric, \textit{derivation coverage}, to generate good samples to drive synthesis. Intuitively, derivation coverage attempts to capture distinct behaviors due to the presence/absence of each of the semantic actions corresponding to the syntax-directed evaluation of different strings. Hence, a directed example generation strategy on example coverage yields smaller example suites that reduce the synthesis time. We omit the details of this coverage metric for want of space (we refer the interested readers to \cite{archivePanini} for the details).

We also propose a new coverage metric, \textit{derivation coverage}, to generate good samples to drive synthesis. Let us explain derivation coverage with an example, \fexs{x\textasciicircum 2+4*x+5}{3} from the grammar in~\figref{lst:motivationGrammar}. The leftmost derivation of this string covers eight productions, (\{1, 2, 4, 5, 6, 7, 10, 11\}) out of a total of 11 productions. Intuitively, it implies that the Hoare triple constraint from its semantic evaluation will \textit{test} the semantic actions corresponding to these productions. 

Similarly, the Hoare logic constraint from the example \fexs{x\textasciicircum 2+7*x+sin(x)}{2} will cover 9 of the productions, \{1, 2, 4, 5, 6, 7, 8, 10, 11\}. As it \textit{also} covers the semantic action for the production 8, it tests an additional behavior of the attribute grammar. On the other hand, the example \fexs{x\textasciicircum 3+5x}{5} invokes the productions, \{1, 2, 4, 5, 6, 7, 10, 11\}. As all these semantic actions have already been \textit{covered} by the example \fexs{x\textasciicircum 2+4*x+5}{3}, it does not include the semantic action of any new set of productions. 

In summary, derivation coverage attempts to abstract the derivation of a string as the \textit{set of productions in its leftmost derivation}. It provides an effective metric for quantifying the quality of an example suite and also for building an effective example generation system.

\noindent {\bf Validation. }%\label{sec:motiValid}}
Our example generation strategy can start off by \textit{sampling} strings $w$ from the grammar (that improve derivation coverage), and context $\beta$; next, it can query the oracle for the intended semantic value $v = Oracle(w\B{\beta})$ to create an example $\semgb{w}=v$.

Consider that the algorithm finds automatically an example \fexs{x\textasciicircum 2+4*x+5}{3}. Now, there are two possible, semantically distinct definitions that satisfy the above constraint (see \figref{fig:mulDef} and \figref{fig:wrongMulDef}), indicating that the problem is underconstrained. Hence, our system needs to select additional examples to resolve this. One solution is to sample multiple contexts on the same string to create multiple constraints:

\begin{itemize}[noitemsep,nolistsep]
    \item $\mathtt{\{x=2\}\  K_1.val \gets 2 + 1\varepsilon; \ldots \{output = 13+6\varepsilon\} }$
    \item $\mathtt{\{x=4\}\  K_1.val \gets 4 + 1\varepsilon; \ldots \{output = 29+10\varepsilon\} }$
\end{itemize}

The above constraints resolve the ambiguity and allows the induction of a semantically unique definitions. The check for semantic uniqueness can be framed as a check for \textit{distinguishing inputs}: given a set of synthesized completion $\gcomplete{{f_1}, \dots, {f_n}}$, we attempt to synthesize an alternate completion $\gcomplete{{g_1}, \dots, {g_n}}$ and an example string $w$ (and context $\beta$) such that $\semb{w}^{\gcomplete{{f_1}, \dots, {f_n}}} \neq \semb{w}^{\gcomplete{{g_1}, \dots, {g_n}}}$. In other words, for the same string (and context), the attribute grammar returns different evaluations corresponding to the two completions. For example, \fexs{x\textasciicircum 2+4*x+5}{2} is a \textit{distinguishing inputs} witnessing the ambiguity between the definitions shown in \figref{fig:mulDef} and \figref{fig:wrongMulDef}.

On the other hand, the algorithm could also have sampled other strings (instead of contexts) for additional constraints. {\tool} prefers the latter; that is, it first generates a \textit{good} example suite (in terms of derivation coverage) and only uses distinguishing input as a validation (post) pass. If such inputs are found, additional contexts are added to resolve the ambiguity.

\forFMCAD{
    We provide the detailed algorithm of example generation in the extended version~\cite{archivePanini}.
}

	%{ \ttfamily NUM : x \textasciicircum\ NUM, NUM : x + x, NUM : x - x, NUM : x * x,\\ NUM : sin(x\textasciicircum 2), NUM : cos(x\textasciicircum 2), NUM : x * cos(x) } \todo{\pankaj{can we do better to show the examples}}

\section{Algorithm}

%; \tool attempts to synthesize the required functionality to achieve behavior specified by the example suite.

Given an attribute grammar $\ghole$, a set of holes $h_i \in H$, a domain-specific language $D$, an example suite $E$ and a context $\beta$, {\tool} attempts to find instantiations $g_i$ for $h_i$ such that,
%\underset{i=1: |H|}
%\[
\begin{equation}
{\bf Find} \{g_1, \dots, g_{|H|}\} \in D \textbf{ such that } \forall \s{s,\beta,v} \in E.\ {\semgb{s}} = v
\label{Eq:synthCons}
\end{equation} %\] 
 \textrm{ where } the attribute grammar $\g=\ghole[g_1/h_1,\dots, g_{|H|}/h_{|H|}]$ and variable bindings $\beta$ maps variables in $s$ to values.

\subsection{Basic Scheme: \textsc{AllAtOnce} \label{sec:allatonce}}
%\begin{wrapfigure}{r}{.54\textwidth}
%	\vspace*{-4.5em}
%	\begin{minipage}{.54\textwidth}
		\begin{algorithm}[t]
			$\varphi \gets \top$\; \label{algo1:phiTrue}
			$\ghole_1 \gets \ghole[R]$\; \label{algo1:gR}
			\For{$\s{w, v} \in T$}{\label{algo1:loop}
				$t \gets \textsc{GenTrace}(\semgho{w})$\; \label{algo1:genTrace}
				$\varphi \gets \varphi \land (out(t) = v)$\;\label{algo1:addCons}
			}
			$B \gets \textsc{Synthesize}(\varphi, D)$\; \label{algo1:synth}
			\Return{B}\;
			\caption{$\textsc{SynthHoles}(\hole{G}, T, R, D)$\label{algo:synthHoles}}
		\end{algorithm}
%	\end{minipage}\vspace*{-3em}
%\end{wrapfigure}

Our core synthesis procedure (Algorithm~\ref{algo:synthHoles}), \textsc{SynthHoles}($\ghole, E, R,$ $D$), accepts a sketch $\ghole$, an example (or test) suite $E$, a set of \textit{ready} functions $R$ and a DSL $D$; all holes whose definitions are available are referred to as \textit{ready} functions. When \textsc{SynthHoles} is used as a top-level procedure (as in the current case), $R=\emptyset$; if not empty, the definitions of the ready functions are substituted in the sketch $\ghole$ to create a new sketch $\ghole_1$ on the remaining holes (Line~\ref{algo1:gR}). We refer to the algorithm where $R=\emptyset$ at initialization as the \textsc{AllAtOnce} algorithm.

Our algorithm exploits the fact that a syntax-directed semantic evaluation of a string $w$ on an attribute grammar $\g$ produces a loop-free program. It attempts to compute a symbolic encoding of this program trace in the formula $\varphi$ (initialized to \texttt{true} in Line~\ref{algo1:phiTrue}). \textsc{GenTrace}() instruments the semantic evaluation on the string $w$ to collect a symbolic trace (the loop-free program) consisting of the set of instructions encountered during the syntax-directed execution of the attribute grammar (Line~\ref{algo1:genTrace}); an output from an operation that is currently a \textit{hole} is appended as a symbolic variable. The assertion that the expected output $v$ matches the final symbolic output $out(t)$ from the trace $t$ is appended to the list of constraints (Line~\ref{algo1:addCons}). Finally, we use a program synthesis procedure, \textit{Synthesize} with the constraints $\varphi$ in an attempt to synthesize suitable function definitions for the holes in $\varphi$ (Line~\ref{algo1:synth}). Given a constraint in terms of a set of input vector $\vec{x}$ and function symbols (corresponding to holes) $\vec{h}$, 
\begin{equation}
\textsc{Synthesize}(\varphi(\vec{x},\vec{h})) := \vec{h} \text{ such that } \exists \vec{h}.\ \forall{\vec{x}}.\ \varphi(\vec{x},\vec{h})\label{Eq:synthesizeFunction}
\end{equation}
% \todo{complete it} \todo{add line number information}
We will use an example from forward differentiation (\figref{lst:motivationGrammar}) to illustrate this. Let us consider two inputs, \texttt{\fex{x\textasciicircum 2-2*x}{3}{3 + 4$\varepsilon$}} and\\ \texttt{\fex{3*x+6}{2}{12+3$\varepsilon$}}. For the first input \texttt{\fex{3*x+6}{2}{12 + 3$\varepsilon$}}, the procedure \textsc{GenTrace()} (Line~\ref{algo1:genTrace}) generates a symbolic trace (denoted $t_1$):
\[\{x_1 = 2+1\varepsilon;\ \alpha_1 = \hole{h_3}(3+0\varepsilon, x_1);\ out_{t1} = \hole{h_1}(\alpha_1, 6+0\varepsilon);\} \]
The following symbolic constraint is generated from above trace:
\[\varphi_{t1} \equiv ( x_1 = 2+1\varepsilon \land \alpha_1 = \hole{h_3}(3+0\varepsilon, x_1) \land out_{t1} = \hole{h_1}(\alpha_1, 6))\]
In the trace $t_1$, operations $\hole{h_1}$ and $\hole{h_3}$ are holes and variables, i.e., $\alpha_1, out_{t1}$ are the fresh symbolic variables. In the next step (line~\ref{algo1:addCons}), the constraints generated from trace $t_1$ is added,
\[\varphi \equiv \top \land (\varphi_{t1} \land out_{t1} = 12 + 3\varepsilon)\]
In the next iteration of the loop at line~\ref{algo1:loop}, the algorithm will take the second input, (i.e., \texttt{\fex{x\textasciicircum 2-2*x}{3}{3 + 4$\varepsilon$}}). In this case, \textsc{GenTrace()} will generate following trace ($t_2$),

\begin{align*}
\{x_2 = 3+1\varepsilon; \alpha_2 = \hole{h_4}(x_2, 2);\ \alpha_3 = \hole{h_3}(2+0\varepsilon, x_2);\\ out_{t2} = \hole{h_2}(\alpha_3, \alpha_4)\}
\end{align*}

The generated constraints from $t_2$ will be, 
\begin{align*}
\varphi_{t2} \equiv &(x_2 = 3+1\varepsilon \land \alpha_2 = \hole{h_4}(x_2, 2) \land \alpha_3 = \hole{h_3}(2+0\varepsilon, x_2) \\ & \land out_{t2} = \hole{h_2}(\alpha_3, \alpha_4))
\end{align*}
At line~\ref{algo1:addCons}, new constraints will be,
\[\varphi \gets \top \land (\varphi_{t1} \land out_{t1} = 12 + 3\varepsilon) \land (\varphi_{t2} \land out_{t2} = 3+4\varepsilon)\]
At line~\ref{algo1:synth}, with $\varphi$ as constraints, the algorithm will attempt to synthesize definitions for the holes (i.e., $\hole{h_1}, \hole{h_2}, \hole{h_3}$ and $\hole{h_4}$). 
%For example synthesized $\hole{add}$ is as follows,
%\begin{align*}
%	\hole{add} =& \lambda \mathtt{a_1, a_2: \{(a_1.real + a_2.real), (a_1.dual + a_2.dual )\}} 
%\end{align*}

%\begin{lstlisting}[escapeinside={@}{@}]
%add(arg1, arg2):
%	return @$\langle$ real(arg1) + real(arg2))@, 
%		 @(dual(arg1) + dual(arg2)) $\rangle$@
%\end{lstlisting}

\subsection{Incremental Synthesis \label{sec:increSynthesis}}
\begin{algorithm}[t]
	%\footnotesize
	$T \gets \emptyset$\;\label{algo2:Rphi}
	%$R \gets \{ (p_i:\{\dots, h_i:\bot, \dots\}) \ |\ p_i \in P, h_i \in H\}$\;
	$R \gets \emptyset$\;\label{algo2:Tphi}
	\While{$T \neq E$}{
		$\s{w,v} \gets \textsc{SelectExample}(E \setminus T)$\;\label{algo2:selectEx}
		$Z \gets \textsc{GetSketchyProds}(\ghole, w)$\;\label{algo2:getSketchy}
		\eIf{$Z \subseteq R$}{ \label{algo2:checkEval}
			$\ghole_1 \gets \ghole[R]$\; \label{algo2:g1gR}
			\eIf{$\semgho{w} = v$}{ \label{algo2:checkRes}
				$T \gets T \cup \{\s{w,v}\}$\;\label{algo2:testPassed}
				continue\;
			}{
				$R \gets R \setminus Z$\;\label{algo2:testFailed}
				$T_e \gets T \cup \{\s{w,v}\}$\;\label{algo2:updatet}
			}
		}{
			$T_e \gets \{\s{w_i, v_i} \ | \ w \cong w_i, \s{w_i,v_i}\in E\}$\;\label{algo2:getTe}
		}
		$B \gets \textsc{SynthHoles}(\ghole, T_e, R, D)$\; \label{algo2:synthProc}
		\If{$B = \emptyset$}{
			\If{$R \cap Z \neq \emptyset$}{ \label{algo2:RandZ}
				$R \gets R \setminus Z$\; \label{algo2:RminusZ}
				$B \gets \textsc{SynthHoles}(\ghole, T \cup T_e, R, D)$\;\label{algo2:reSynth}
			}
			\If{$B = \emptyset$}{
				\Return{$\emptyset$}\; \label{algo2:returnPhi}
			}
		}
		$R_{f} \gets \{ (p_i:\{\dots, h_i\rightarrow B[h_i], \dots\}) \ |$\newline
		\hspace*{2cm}$p_i \in Z \setminus R, h_i \in holes(\Gamma(p_i)) \}$\;
		%\tcc{\footnotesize $B \equiv\{ (p_i:\{\dots, h_i\rightarrow g_i, \dots\}) \ |\ p_i, h_i \in \varphi\}$}
		$R \gets R \cup R_{f}$\;
		$T \gets T \cup T_e$\;
	}
	\Return{$R$}\;	
	\caption{\textsc{SynthAttrGrammar}($\ghole$, E, D)\label{algo:atrgrmsynth}}
\end{algorithm}
The \textsc{AllAtOnce} algorithm exhibits poor scalability with respect to the size of the grammar and the number of examples. The route to a scalable algorithm could be to incrementally learn the definitions corresponding to the holes and make use of the functions synthesized in the previous steps to discover new ones in the subsequent steps. 

However, driving synthesis one example at a time will lead to overfitting. We handle this complexity with a two-pronged strategy: (1) we partition the set of examples by the holes for which they need to synthesize actions, (2) we solve the synthesis problems by their difficulty (in terms of the number of functions to be synthesized) that allows us to memoize their results for the more challenging examples. We refer to this example as the \textsc{IncrementalSynthesis} algorithm.

\p{Derivation Congruence} We define an equivalence relation, \textit{derivation congruence}, on the set of strings $w \in \mathcal{L(G)}$: strings $w_1$, $w_2 \in \mathcal{L(G)}$ are said to be \textit{derivation congruent}, $w_1 \cong_G w_2$ w.r.t. the grammar $G$, if and only if both the strings $w_1$ and $w_2$ contain the same set of productions in their respective derivations. %In other words, both the strings generate the same parse tree on the given grammar.
For example, $w_1:$\fexs{3*x+5}{2}, $w_2:$\fexs{5*x+12}{3} and $w_3:$\fexs{4*x+7*x}{3}.

Note that though the strings $w_1$, $w_2$ and $w_3$ are derivation congruent to each other, while $w_1$ and $w_2$ have similar parse trees, $w_3$ has a quite different parse tree. So, intuitively, all these strings are definition congruent to each other, as, even with different parse trees, they involve the same set of productions (\{1, 2, 4, 5, 6, 10, 11\}) in their leftmost derivation.

Algorithm~\ref{algo:atrgrmsynth} shows our incremental synthesis strategy. Our algorithm maintains a set of examples (or tests) $T$ (line~\ref{algo2:Tphi}) that are consistent with the current set of synthesized functions for the holes; the currently synthesized functions (referred to as \textit{ready} functions), along with the respective ready productions, are recorded in $R$ (line~\ref{algo2:Rphi}). The algorithm starts off by selecting the \textit{easiest} example $\s{w,v}$ at line~\ref{algo2:selectEx} such that the cardinality of the set of sketchy production, $Z$, in the derivation of $w$ is the minimal among all examples not in $T$. The set $R$ maintains a map from the set of sketchy productions to a set of assignments to functions synthesized (instantiations) for each hole contained in the respective semantic actions. %We overload set operations for $R$ to operate on the set of sketchy productions.\todo{not clear}

If all sketchy productions, $Z$, in the derivation of $w$ are now \textit{ready}, we simply \textit{test} (line~\ref{algo2:checkEval}) to check if a syntax-guided evaluation with the currently synthesized functions in $R$ yield the expected value $v$: if the test passes, we add the new example to the set of passing examples in $T$ (line~\ref{algo2:testPassed}). Otherwise, as the current hole instantiations in $R$ is not consistent for the derivation of $w$, at line~\ref{algo2:testFailed} we remove all the synthesized functions participating in syntax-directed evaluation of $w$ (which is exactly $Z$). Furthermore, removal of some functions from $R$ requires us to re-assert the new functions on all the past examples (contained in $T$) in addition to the present example (line~\ref{algo2:updatet}). 

If all the sketchy productions ($Z$) in the derivation of $w$ are not ready, we attempt to synthesize functions for the \textit{missing} holes, \textit{with the set of current definitions in $R$ provided in the synthesis constraint}. 

The synthesis procedure (line~\ref{algo2:synthProc}), if successful, yields a set of function instantiation for the holes. In this case, the solution set from $B$ is accumulated in $R$, and the set of \textit{passing} tests extended to contain the new examples in $T_e$.

However, synthesis may fail as some of the current definitions in $R$ that were assumed to be correct and included in the synthesis constraint is not consistent with the new examples ($T_e$). In this case, we remove the instantiations of all such holes occurring in the syntax-directed evaluation of $w$ (line~\ref{algo2:RminusZ}) and re-attempt synthesis (line~\ref{algo2:reSynth}). If this attempt fails too, it implies that no instantiation of the holes exist in the provided domain-specific language $D$ (line~\ref{algo2:returnPhi}).%\todo{add line number information}

\forFMCAD{We provide a detailed example on the run of this algorithm in the extended version~\cite{archivePanini}.}

\forarxiv{
\noindent {\bf Example.} We will use the following example set $E$ to explain the Algorithm~\ref{algo:atrgrmsynth}: \\
$E$ : \Big\{  \fex{cos(x\textasciicircum 2)}{2}{-0.65 + 3.02$\varepsilon$},\\ \fex{x*x}{4}{16 + 8$\varepsilon$}, \\\fex{x*cos(x)}{4}{-2.61 + 2.37$\varepsilon$}\Big\}

The algorithm starts with two empty sets $T$ and $R$. At line~\ref{algo2:selectEx}, \textsc{SelectExample}() selects the example \fex{x*x}{4}{16 + 8$\varepsilon$}; here $w$ is \fexs{x*x}{4} and $v$ is \texttt{16 + 8$\varepsilon$}.
At line~\ref{algo2:getSketchy}, \{$\hole{h_3}$\} is assigned to $Z$, as the semantic evaluation of this example uses the definition of $\hole{h_3}$. Since $R$ is $\emptyset$ at line~\ref{algo2:checkEval}, at line~\ref{algo2:getTe}, $T_e$ is assigned to all derivation congruent examples to the current input $\langle w,v\rangle$.

At line~\ref{algo2:synthProc}, {\tool} calls the synthesis engine to synthesize the \textit{holes} consistent with set $T_e$. The synthesis engine discharges $\hole{h_3}$ as shown in~\figref{fig:wrongMulDef}. Now set $R$ is accumulated in \{$\hole{h_3}$\} (i.e. it is marked \textit{ready}). 

In the second iteration of the loop, at line~\ref{algo2:selectEx}, \textsc{SelectExample}() selects \fex{cos(x\textasciicircum 2)}{2}{-0.65 + 3.02$\varepsilon$} as input; this case, $Z$ is updated to \{$\hole{h_4}, \hole{h_6}$\} (line~\ref{algo2:getSketchy}). Since $R=\{\hole{h_3}\}$, {\tool} execute line~\ref{algo2:getTe} to compute the derivation congruence for the current input string. This set of examples will be appended to $T_e$ which will be used for synthesis. At line~\ref{algo2:synthProc}, \textsc{SynthHoles}() takes the attribute grammar, the set of examples $T_e$ and the ready set $R$ so far i.e., \{$\hole{h_3}$\}  along with DSL. \textsc{SynthHoles}() will try to synthesize the holes present in the $Z$, i.e., \{$\hole{h_4}, \hole{h_6}$\}. Since both holes in $Z$ are unfrozen \textsc{SynthHoles}() will be able to synthesize the definitions for both \{$\hole{h_4}, \hole{h_6}$\}. After synthesizing \{$\hole{h_4}, \hole{h_6}$\} at line~\ref{algo2:synthProc}, {\tool} updates the $R$ to \{$\hole{h_3}, \hole{h_4}, \hole{h_6} $\}.

In the third iteration, with \fex{x*cos(x)}{4}{-2.61 + 2.37$\varepsilon$} as input, line~\ref{algo2:getSketchy} updates $Z$ to \{$\hole{h_3}, \hole{h_6}$\}. Since $Z \subseteq R$ at line~\ref{algo2:checkEval}, we try to evaluate the expression \fexs{x*cos(x)}{4} using the available definition of $\hole{h_3}, \hole{h_6}$. At line~\ref{algo2:g1gR} we create another attribute grammar $\ghole_1$ which contains the definitions present in $R$. At line~\ref{algo2:checkRes} the output result of input \fexs{x*cos(x)}{4} using definitions present in $R$ does not match the solution \texttt{-2.61 + 2.37$\varepsilon$} provided by oracle. So we can conclude that the definitions present in the $R$ (\{$\hole{h_3}, \hole{h_6}$\}) is wrong. In this if we inspect manually synthesized version of $\hole{h_3}$ (\figref{fig:wrongMulDef}) is incorrect, but {\tool} has no information to detect this, hence {\tool} removes \{$\hole{h_3}, \hole{h_6}$\} from \textit{ready} set $R$ at line~\ref{algo2:testFailed}.
%However, definition of $\hole{h_3}$ (\figref{fig:wrongMulDef}) is incorrect, ???\todo{how do we know it is incorrect?} so we will remove \{$\hole{h_3}, \hole{h_6}$\} ($Z$), from $R$ at line~\ref{algo2:testFailed}.
At line~\ref{algo2:synthProc}, {\tool} will call \textsc{SynthHoles}() with \{$\hole{h_4}$\} as $R$ and $T\cup\fexs{x*cos(x)}{4}$ as $T_e$. Synthesized correct definition of $\hole{h_3}$ and $\hole{h_6}$ as shown in~\figref{fig:cosDef} and \figref{fig:mulDef}, respectively.
Since, $E$ is empty (i.e. all examples have been processed), {\tool} returns the definitions of $\hole{h_3}, \hole{h_4}$ and $\hole{h_6}$.

Let us consider another scenario: in second iteration of the loop, \textsc{SelectExample}() selects \fex{x*cos(x)}{4}{-2.61 + 2.37$\varepsilon$} (instead of \\{\fex{cos(x\textasciicircum 2)}{2}{-0.65 + 3.02$\varepsilon$}}). Now, at line~\ref{algo2:getSketchy}, $Z$ is updated to \{$\hole{h_3}, \hole{h_6}$\}. At this point, after first iteration with input \fex{x*x}{4}{16 + 8$\varepsilon$} $R$ contains \{$\hole{h_3}$\}. 
%hence {\tool} executes\todo{what happens here? explain clearly} line~\ref{algo2:getTe}. Since, $\hole{h_3}$ is synthesized incorrectly, so 
At line~\ref{algo2:synthProc} the synthesis procedure is being called with given $R$ and the input examples $T_e$ collected from the derivation coverage to synthesize \{$\hole{h_3}, \hole{h_6}$\}. This call to \textsc{SynthHoles} returns unsat due to the wrong definition of $\hole{h_3}$. Due to failure {\tool} tries to synthesize again by updating $R$ with removal of $\hole{h_3}$ (content of $Z$) from $R$ to discard the available definition of $\hole{h_3}$. At line~\ref{algo2:reSynth} {\tool} reattempts synthesis with both holes, $\hole{h_3}$ and $\hole{h_6}$. Since this time both holes  \{$\hole{h_3}, \hole{h_6}$\} are unfrozen hence \textsc{SynthHoles} succeed in the synthesis their correct definitions (\figref{fig:mulDef} and \figref{fig:cosDef}). 

With input \fex{cos(x\textasciicircum 2)}{2}{-0.65 + 3.02$\varepsilon$} in the next iteration, {\tool} will make $\hole{h_4}$ as hole as definition of $\hole{h_6}$ is already present. As the input set is empty {\tool} will terminated after returning synthesized definitions of $\hole{h_3}, \hole{h_4}$ and~$\hole{h_6}$.

\subsection{Automated Example Generation}
\begin{algorithm}[t]
	%    \textbf{global } $Q_1 = \emptyset$\;
	%    \textbf{global } $E = \emptyset$\;
	%\textbf{global } $AllCoverage = \emptyset$\;
	\If{$X \in T$}{
		\Return{concat(w, X)}\; %\tcc{$w$ passed by reference}
	}
	$C \gets \{p_i \ |\ head(p_i) = X\}$\;
	p $\gets \textsc{Select}(C)$\;
	$deriv \gets deriv \cup \{p\}$\;
	\For{$Y \in body(p)$}{
		$\textsc{GenerateExample}(\ghole, Y, Oracle, deriv, w, Q_1\})$\; 
	}
	\eIf{$X=S \land deriv \not\subseteq Q_1$}{
		\Return{$deriv, \s{w, Oracle(w)}$}\;
	}{
		\Return{$\bot$}\;
	}
	\caption{\textsc{GenerateExample}($\ghole, X, Oracle,$\\ \hspace*{5cm} $\textbf{ref}\ deriv, \textbf{ref}\ w, Q_1$)\label{algo:genex}}
\end{algorithm}
%An \textit{example suite}, E, is a set of partition of examples \{$P_0$, \dots, $P_i$, \dots, $P_n$\}, such that for any $s_a, s_b \in P_i$ implies $s_a \cong s_b$.
Given an input-output oracle, \textit{Oracle}, we also provide an automated strategy to generate examples. For the generation of effective examples, we define a new coverage metric on grammars, that we refer to as \textit{derivation coverage}.

\p{Production Set} Given a string $w \in \mathcal{L(G)}$, we define the \textit{production set} ($ProdSet(w)$) of $w$ as the set of productions occurring in the derivation of $w$.

\p{Derivation Coverage} Given a context-free grammar $\g$, we define the derivation coverage of a set of strings $W=\{w_1, w_2, \dots, w_n\}$ as the fraction of the production sets of the derivations of the strings in $W$ with respect to all possible sets of productions. More formally, the derivation coverage of $W$ is $\frac{|Q_1|}{|Q_2|}$, where %\todo{needs space}

\begin{itemize}[noitemsep,nolistsep]
	\item $Q_1 = \{ ProdSet(w_k) \ | w_k \in W \}$
	\item $Q_2 = 2^{|P|}$, where $P$ is the set of productions in $\g$
\end{itemize}

Intuitively, derivation coverage attempts to capture distinct behaviors due to the presence/absence of each of the semantic actions corresponding to the syntax-directed evaluation of different strings.

Algorithm \ref{algo:genex} shows the core of our test generation strategy; when invoked as $\textsc{GenerateExample}(\ghole, \mathsf{S}, Oracle, deriv, w, Q_1)$, (with $deriv = \emptyset$ and $w=``"$ passed by reference). In case there are multiple productions whose head is $X$, our procedure non-deterministically selects a production using the \textsc{Select()} routine.
%\textsc{Select()} is a  non-deterministic selection routine, which non-deterministically choose a production from the set of productions. 
\textsc{GenerateExample}() returns a test only if it increases derivation coverage (i.e., if the cardinality of $Q_1$ increases). The test generator makes multiple calls to \textsc{GenerateExample}() till an adequate number of examples are generated (or a desired coverage bound is reached). Our engine also attempts to prioritize smaller derivations and smaller examples. 
 
}

\p{Theorem} If the algorithm terminates with a non-empty set of functions, $\ghole$ instantiated with the synthesized functions will satisfy the examples in $E$; that is, the synthesized functions satisfy Equation~\ref{Eq:synthCons}.

The proof is a straightforward argument with the inductive invariant that at each iteration of the loop, the $\ghole$ instantiated with the functions in $R$ satisfy the examples in $T$.

%\section{Automatically generating examples}

\section{Experiments}

Our experiments were conducted in Intel(R) Xeon(R) CPU E5-2620 @ 2.00GHz with 32 GB RAM and 24 cores on a set of benchmarks shown in Table~\ref{tab:desc}. {\tool} uses \textsc{Flex}~\cite{flex} and \textsc{Bison}~\cite{bison} for performing a syntax-directed semantic evaluation over the language strings.
{\tool} uses \textsc{Sketch}~\cite{sketch} to synthesize function definitions over loop-free programs, and the symbolic execution engine \textsc{Crest}~\cite{crest} for generating example-suites guided by derivation coverage.

\begin{figure*}[t]
				\footnotesize
	\begin{subfigure}{.3\textwidth}
	\centering
	\footnotesize
	%	\begin{framed}
	\begin{tabular}{>{\color{red}}l >{\color{blue}}l}
		%		S \hspace*{1.5mm}$\rightarrowtail$  INCLUDE \textless\ H \textgreater\ MAIN & \\
		%		\hspace*{5mm}\ $\mid$\ \ INCLUDE `` H " MAIN&\\
		S $\rightarrowtail$ MAIN B&\\
		\hspace*{6mm}$\mid$\ MAIN B&\\
		
		\ldots&\\
		A \hspace*{1mm}$\rightarrowtail$  T&\\
		\hspace*{6mm}$\mid$\ \ \ E +  T & {A.val = \highlight{\hole{h_a}(E.val, T.val)};}\\
		\hspace*{6mm}$\mid$\ \ \  E - T & {A.val = \highlight{\hole{h_b}(E.val, T.val)};}\\
		
		T\hspace*{2.5mm}$\rightarrowtail$ F&\\
		\hspace*{6mm}$\mid$\ \ \ T * F & {T.val = \highlight{\hole{h_c}(T.val, F.val)};}\\
		\hspace*{6mm}$\mid$\ \ \ T / F & {T.val = \highlight{\hole{h_d}(T.val, F.val)};}\\
		\ldots &\\
	\end{tabular}
	%	\end{framed}
	\caption{Attribute grammar sketch for mini-compiler\label{fig:miniCodeSketch}}
	\end{subfigure}	
	\hfill
	\begin{subfigure}{.2\textwidth}
		%				\hspace*{2em}
		\begin{lstlisting}[mathescape=true]
int main{
  int a, b, c;
  a = 4;
  b = a + 3; //$\hhole{h_a}$
  c = a - b; //$\hhole{h_b}$
  return c;
}
		\end{lstlisting}
		\caption{A simple \texttt{C} code \label{fig:miniCode}}
	\end{subfigure}
	\hfill
	\hspace*{.5cm}
	\begin{subfigure}{.2\textwidth}
					\begin{tabular}{|p{20pt} p{10pt} p{10pt} p{10pt}|}
			\hline
									$op$ &   ${arg_1}$  &     ${arg_2}$    &    $dst$\\
			\hline \hline
			assign         &   4     &            &	   a\\
			$\hhole{h_a}$  &   a     &      9     &    T0\\
			assign         &   T0    &            &	   b\\
			$\hhole{h_b}$  &   a     &      b     &    T1\\
			assign         &   T1    &            &    T2\\
			ret  	       &	     &		      &	   T2\\
			\hline
		\end{tabular}
		\caption{Three-address code generated from \texttt{C} code in~\figref{fig:miniCode} \label{fig:miniQuad}}
	\end{subfigure}	
	\hfill
	\hspace*{.5cm}
	\begin{subfigure}{.2\textwidth}
		\DontPrintSemicolon
		\SetAlgoNoLine
		\SetKwProg{add}{\colorbox{yellow}{$\hole{h_a}$}}{:}{}
		\add{$a$\ $b$}{
			$emit(``load\ r_1\ a")$\\
			$emit(``load\ r_2\ b")$\\
			$emit(``plus\ r_1\  r_2")$\\
		}
		\SetKwProg{sub}{\colorbox{yellow}{$\hole{h_b}$}}{:}{}
		\sub{$a$\ $b$}{
			$emit(``load\ r_1\ a")$\\
			$emit(``load\ r_2\ b")$\\
			$emit(``sub\ r_1\  r_2")$\\
		}
		%	\vspace*{1.5cm}
		\caption{Synthesized definition for $\hole{h_a}, \hole{h_b}$ \label{fig:miniSol}}
	\end{subfigure}	
	\caption{Synthesis of mini-compiler (b12) \label{fig:phasesMini}}
\end{figure*}

\begin{figure*}[t]
	\begin{minipage}{.6\linewidth}
	\footnotesize
					\captionof{table}{Description of benchmarks} \label{tab:desc}
	\centering
	\begin{tabular}{|>\bfseries c|c||c|c|p{2.3cm}||c|c||c|} %unhode to get description
		%   	 \begin{tabular}{|c|p{3cm}|H c|c|c|}
		\hline
		\textbf{Id} & \textbf{Benchmark} & \textbf{\#P} & \textbf{\#H} & \textbf{Example} & \textbf{\#R}  & \multicolumn{2}{c|}{\bf Time (s)}\\
		\cline{7-8}
		 &&&&&&\textbf{AAO}  &  \textbf{IS}\\
		\hline
		\hline
		b1 & Count ones & 5 & 1 & $11001$ & 0 & 3.2  & 3.1\\ 
		\hline
		b2 & Binary to integer & 5 & 1 & $01110$ & 0 & 3.6 & 2.9 \\ 
		\hline
		b3 & Prefix evaluator & 7 & 4 & $+\ 3\ 4$ & 0 &  TO& 10.1 \\
		\hline
		b4 & Postfix evaluator  & 7 & 4 & $2\ 3\ 4\ *\ +$ & 0 & TO & 10.5\\
		\hline
		b5 & Arithmetic calculator & 8 & 4 & $5 * 2 + 8$ & 0 & TO & 12.8 \\
		\hline
		b6 & Currency calculator & 10 & 4 & $\mathtt{USD}\ 3\ +\ \mathtt{INR}\ 8$ & 0 & TO & 13.6 \\
		\hline
		b7 & if-else calculator & 10 & 4 & \texttt{if(3+4 == 3)} \newline \texttt{then  44;} \newline \texttt{else 73;} & 1 & TO & 21.7\\
		\hline
		b8 & Activation record layout & 10 & 3 & \texttt{int a , b;} & 0 & TO & 13.8 \\
		\hline
		b9 & Type checker & 11 & 5 & \texttt{(5 - 2) == 3} & 1 & TO & 15.4 \\
		\hline
		b10 & Forward differentiation & 20 & 12 & \texttt{x*pow}(\texttt{x},$3$) & 2  & TO & 39.2 \\
		\hline
		b11 & Bytecode interpreter & 39 & 36 & \texttt{bipush 3}; \newline \texttt{bipush 4};\newline \texttt{iadd}; & 3 & TO & 141.4 \\
		\hline
		b12 & Mini-compiler & 43 & 6 & \ttfamily int main()\{\newline\ return 2+3;\} & 0 & TO & 9.2\\
		\hline
	\end{tabular}
	\end{minipage}
	\hspace{3em}
	\begin{minipage}{.3\linewidth}
		\centering
		\begin{minipage}{\linewidth}
			\begin{subfigure}[b]{\linewidth}
				
				\begin{lstlisting}[basicstyle=\ttfamily\small]
init {
 run Foo(8+(6-7));
}
				\end{lstlisting}
				\caption{\textsc{Promela} source code\label{fig:promelaAst1}}
			\end{subfigure}
%			\hspace{-.4cm}
			\begin{subfigure}[b]{\linewidth}
				\begin{lstlisting}[basicstyle=\ttfamily\small,escapeinside={(*}{*)}]
node n1 = node(val=8);
node n2 = node(val=6);
node n3 = node(val=7);
node n4 = (*\colorbox{yellow}{$\hole{h_a}$}*)(n2, n3);
node n5 = (*\colorbox{yellow}{$\hole{h_b}$}*)(n1, n4);
node n6 = (*\colorbox{yellow}{$\hole{h_c}$}*)(`Foo',n5);
				\end{lstlisting}
				\caption{Trace generated \label{fig:promelaAst2}}
			\end{subfigure}
			\caption{Trace generation for AST construction		\label{fig:promelaAst}}
		\end{minipage}
%		\begin{figure}

%		\end{figure}
	\end{minipage}
\end{figure*}

\noindent We attempt to answer the following research questions:%\todo{To modify the RQs}
\begin{itemize}[noitemsep,nolistsep]
    \item Can {\tool} synthesize attribute grammars from a variety of sketches? 
    \item How do \textsc{IncrementalSynthesis} and \textsc{AllAtOnce} algorithms compare?
    \item How does {\tool} scale with the number of holes?
    \item How does {\tool} scale with the size of the grammar?
\end{itemize}

    The default algorithm for {\tool} is the \textsc{IncrementalSynthesis} algorithm; unless otherwise mentioned, {\tool} refers to the implementation of \textsc{IncrementalSynthesis} (Algorithm~\ref{algo:atrgrmsynth}) for synthesis using examples generation guided by derivation coverage\forFMCAD{~\cite{archivePanini}} \forarxiv{(Algorithm~\ref{algo:genex})}. While \textsc{AllAtOnce} works well for small grammars with few examples, \textsc{IncrementalSynthesis} scales well even for larger grammars, both with the number of holes and size of the grammar.

    %We have evaluated {\tool} across a set of benchmarks spanning from 5 to 43 productions. 
    
    {\tool} can synthesize semantic-actions across both synthesized and inherited attributes. Some of our benchmarks contain inherited-attributes: for example, benchmark {b8} uses inherited-attributes to pass the type information of the variables. Inherited-attributes pose no additional challenge; they are handled by the standard trick of introducing “marker” non-terminals~\cite{aho-sethi-ulman}.

\subsection{Attribute Grammar Synthesis}

We evaluated {\tool} on a set of attribute grammars adapted from software in open-source repositories~\cite{calcCite, postfixSite, mini-compiler, bin2intSite, forwardDiff, typeCheck, aho-sethi-ulman}. Table~\ref{tab:desc} shows the benchmarks, number of productions (\textbf{\#P}), number of holes (\textbf{\#H}), input example,  solving time  (\textbf{Time}, {\bf AAO} for \textsc{AllAtOnce} and {\bf IS}  for \textsc{IncrementalSynthesis}) and number of times a defined function was refuted (\textbf{\#R}). Please recall that \textsc{AllAtOnce} refers to Algorithm~\ref{algo:synthHoles} (\S\ref{sec:allatonce}) and \textsc{IncrementalSynthesis} refers to Algorithm~\ref{algo:atrgrmsynth} (\S\ref{sec:increSynthesis}).

 %\todo{remove desc and put it itemized with benchmark id} \todo{fix table, multirow heading, Time(s) first row and algo name in second row}

%See in Appendix~\S\ref{app:benchmarks} for descriptions of benchmarks from b1 to b7.

\forFMCAD{We provide more detailed descriptions of these benchmarks in the extended version~\cite{archivePanini}.
	
The benchmark b10 is the forward differentiation example described in \S\ref{sec:moti}. Benchmarks b11 and b12 are quite complex benchmarks that interpret a (subset) of Java bytecode and compile C code:}

\forarxiv{Brief description for benchmarks provided below:}

\begin{enumerate}[label=b{\arabic*},noitemsep,nolistsep]

\forarxiv{	
\item
\p{Count ones}
Counts the number of ones ($1$) in the input bit-vector. Eg. For the bit-vector $(0111)_2$, it returns $3$.

\item
\p{Binary to integer}
Converts an input bit vector to its corresponding integer value. For example, with $(0110)_2$ as input, we will get $6$ as an output.

\item
\p{Prefix evaluator} 
Evaluates a prefix expression. %Eg. For input, $+\ 4\ *\ 5\ 2$, it produces 14 as output.

\item
\p{Postfix evaluator}
Evaluate postfix expressions. %Eg. On $2\ 3\ 4\ *\ +$, it produces $14$ as output.

\item
\p{Arithmetic calculator}
Calculator for arithmetic expressions. %Eg. $3\ +\ 4\ /\ 2\ -\ 5\ *\ 2$

\item
\p{Currency calculator}
It performs arithmetic operations on currency \texttt{USD} and \texttt{EUR}. %\todo{don't do INR, do USD and EUR}.
Each currency is handled as a distinct type: while addition and subtraction of different currencies is allowed, multiplication and division of two distinct currencies is not allowed. But, one can multiply or divide a currency by scalars. For input, $\mathtt{USD}\ 3\ +\ \mathtt{EUR}\ 8$ it emits $\mathtt{USD}\ 12.28$ assuming $\mathtt{EUR}1\ 1\ =\ \mathtt{USD}\ 1.16$. 

\item
\p{if-else calculator}
Arithmetic calculator with support for branching.

\item 
\p{Activation record layout engine}
This grammar accepts a sequence of variable declarations to construct a map of a procedure activation record. Each variable is allocated memory depending on declared type.

\item 
\p{Type checker}
This grammar accepts an expression over integers and booleans; it perform type checking w.r.t a set of typing rules and signals any type error.

\item 
\p{Forward differentiation} See Section~\ref{sec:moti}
}

\setcounter{enumi}{10}
%	\item
%	\p{Forward differentiation\label{sec:forward}}
%	See~\S\ref{sec:moti}. %To implement \texttt{sin} and \texttt{cos} in sketch we used taylor's expansion of both functions and considered upto two terms.
	
	\item
	\p{Bytecode interpreter\label{sec:bytecode}}
	        Interpreter for a subset of Java bytecode; it supports around 36 instructions~\cite{byteCodeWiki} of different type, i.e., load-store, arithmetic, logic and control transfer instructions. %Table~\ref{tab:byteCode} in Appendix~\ref{app:bytecode} shows the different instructions present in this benchmark.
	        %Synthesizing these functions helps us to do repeated and arduous work. If we consider \texttt{istore\_0} and \texttt{istore\_1}, the both int value into variable 0 and 1 respectively. Though they both are store operations but they differ by the variable index. In this case {\tool} helps by synthesizing repetitive works. {\tool} can synthesize all instructions shown in Table~\ref{tab:byteCode} in less than 2 minutes.
	
		\forarxiv{
		Table~\ref{tab:byteCode} shows the different types of instructions present in this benchmark.

		\begin{table}[H]
			\centering
			\caption{Bytecode instructions \label{tab:byteCode}}
			%	\hspace*{-2em}
			\begin{tabular}{|c|c|c|c|}
				
				\hline
				istore  & istore\_0 & istore\_1 & istore\_2 \\
				
				istore\_3 & iload & iload\_0  & iload\_1 \\
				
				iload\_2 & istore\_3 & iconst\_m1 & iconst\_4 \\
				
				iconst\_5  & iconst\_0 & iconst\_1 & iconst\_2 \\
				
				iconst\_3 & bipush  & ifeq  & ifne \\
				
				iflt & ifle & ifgt & ifge \\  
				
				if\_icmpge & if\_icmple & if\_icmpeq & if\_icmplt \\
				
				if\_icmpgt & iinc &	iadd  & isub \\
				
				imul & idiv & irem & ineg \\
				\hline	
			\end{tabular}
			
		\end{table}
		}
		
	\item
	\p{Mini-compiler\label{sec:miniCom}}
%	        Compiles a subset of \texttt{C} language~\cite{mini-compiler} to two-address intermediate code. Appendix~\ref{app:mini} shows how a simple \texttt{C} program is compiled into two-address code. 	        
%	         In \figref{fig:miniSol} we show two of our synthesized semantic actions.  Here, we use the target language as our DSL that describes the two-address code to generate on encountering certain constructs in the source.
\figref{fig:phasesMini} shows the different steps of synthesizing semantic actions in mini-compiler. \figref{fig:miniCode} is a sample input for the mini-compiler. \figref{fig:miniCodeSketch} shows snippet of the attribute grammar for mini-compiler. \figref{fig:miniQuad} shows the two-address code generated from the input code shown in \figref{fig:miniCode}, where $\hole{h_a}$ and $\hole{h_b}$ are two holes in the two-address code. Finally, in \figref{fig:miniSol} shows the synthesized definition for $\hole{h_a}$ and $\hole{h_b}$ in the target language for two-address code.

\end{enumerate}

\begin{figure}
	\centering
	\includegraphics[scale=.5]{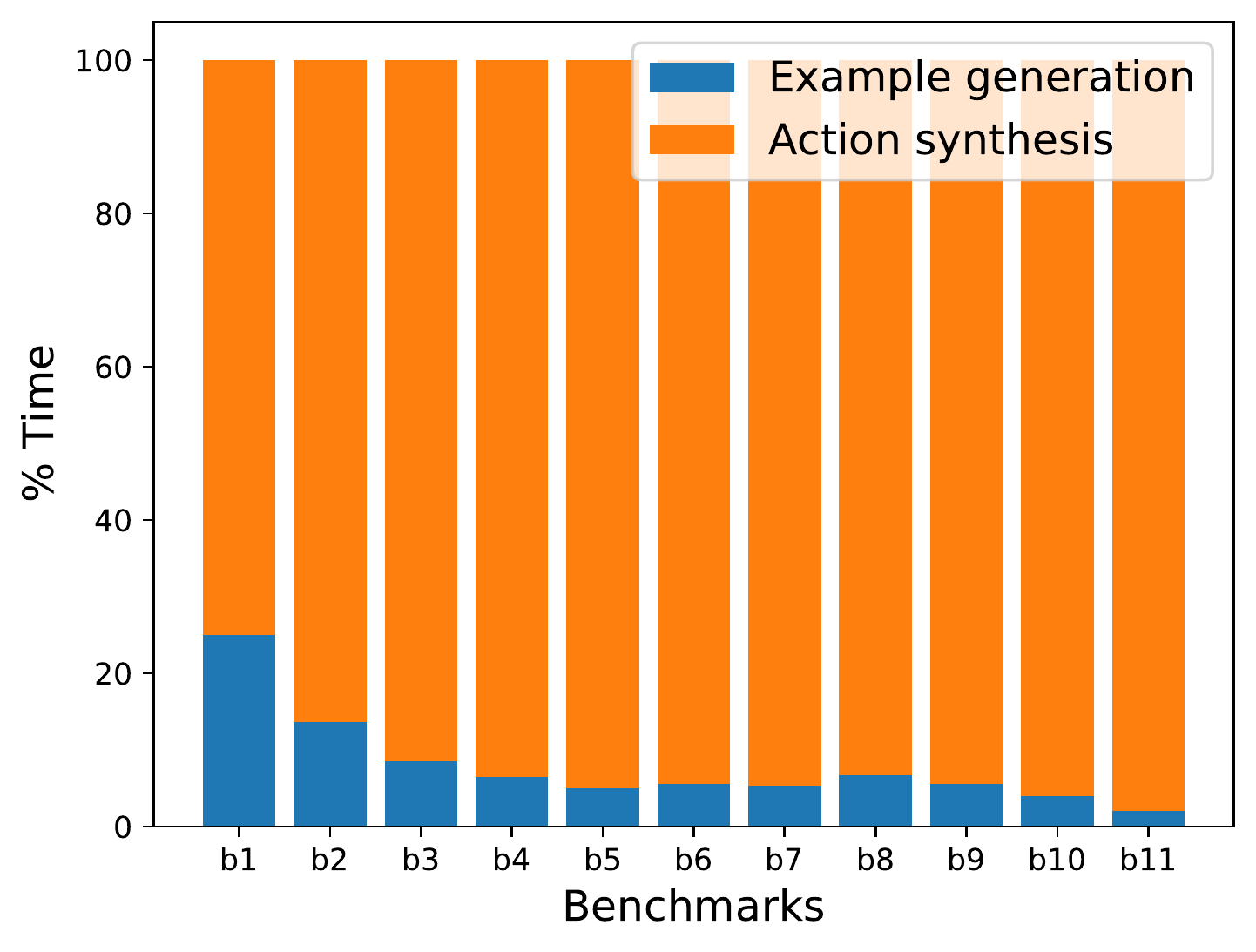}
	\captionof{figure}{\small Stacked bar graph for the \% time spent in example creation and synthesis\label{fig:stack}}
\end{figure}

\figref{fig:stack} attempts to capture the fraction of time taken by the different phases of {\tool}: example generation and synthesis. Not surprisingly, the synthesis phase dominates the cost as it requires several invocation of the synthesis engines, whereas, the example generation phase does not invoke synthesis engines or smt solvers. Further, the difference in time spend in these two phases increases as the benchmarks get more challenging.

\subsection{\textsc{AllAtOnce} v/s \textsc{IncrementalSynthesis}}
\subsubsection{Scaling with holes}
%\todo{TODO}
% Line plot of all benchmarks in same graph with different colors, x-axis time, y-axis \# holes

\begin{figure*}[t]
%	\begin{tabular}{lr}
%\begin{minipage}{.5\linewidth}
		\begin{minipage}{0.55\linewidth}
%		\centering
		\begin{subfigure}[b]{.49\textwidth}
			\includegraphics[scale=0.4]{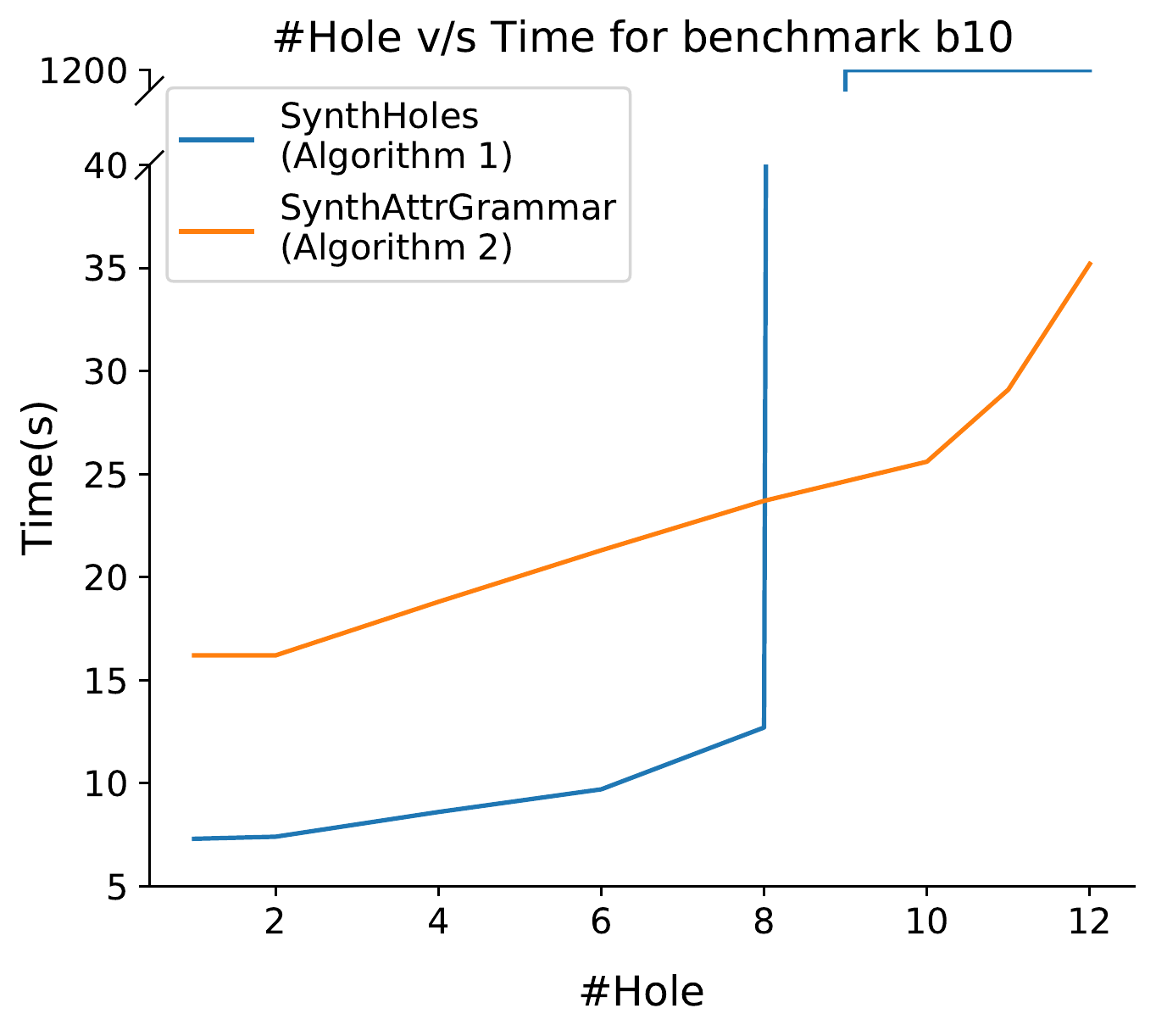}
			\caption{Forward differentiation (b10)\label{fig:holeTimeFD}}
		\end{subfigure}
		\begin{subfigure}[b]{.49\textwidth}
			\includegraphics[scale=0.4]{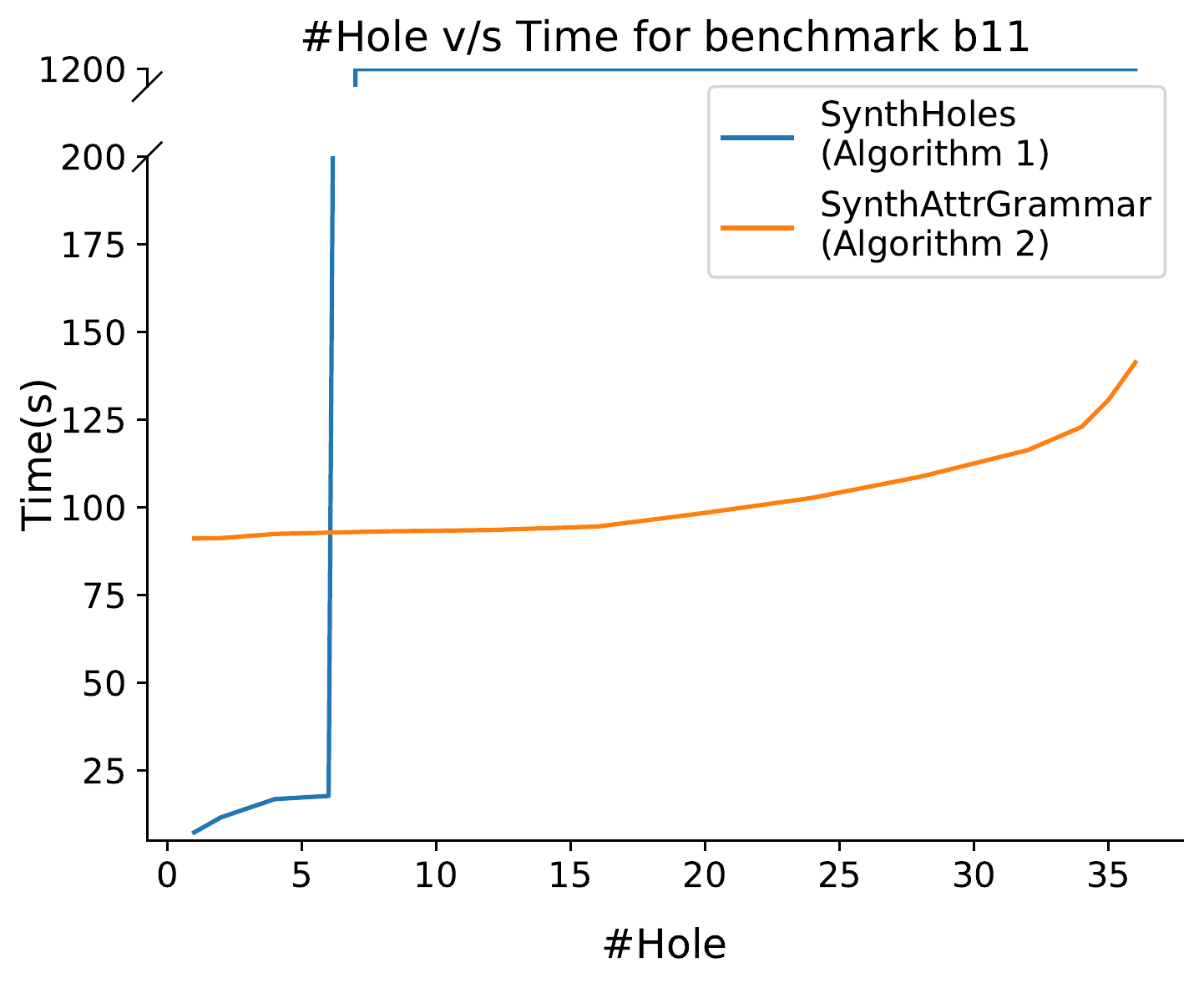}
			\caption{Java bytecode (b11)\label{fig:holeTimeJB}}
		\end{subfigure}
		\caption{\#Hole v/s Time for benchmarks b10 and b11}
		\end{minipage}
%	\end{minipage}
%		&	
%\hspace{.6cm}
\begin{minipage}{.5\linewidth}
	\begin{subfigure}{\linewidth}
	\hspace{-1cm}
	\begin{lstlisting}[basicstyle=\ttfamily\small]
	init {
	int flags[(5 * 25) - 42];
	int v = flags[10 - 4 + (9 / 3)]; 
	}
	\end{lstlisting}
	\caption{\textsc{Promela} source \label{fig:constFold2}}
\end{subfigure}
\hspace{2em}
\begin{subfigure}{\linewidth}
	\begin{lstlisting}[basicstyle=\ttfamily\small]
	init {
	int flags[83];
	int v = flags[9]; 
	}
	\end{lstlisting}
	\caption{\textsc{Promela} optimized  \label{fig:constFold1}}
\end{subfigure}
\caption{Constant folding in \textsc{Promela} \label{fig:constFold}}
\end{minipage}
\end{figure*}

%\begin{table}[H]
%	\centering
%	\caption{Bytecode instructions \label{tab:byteCode}}
%	%	\hspace*{-2em}
%	\begin{tabular}{|c|c|c|c|}
%		
%		\hline
%		istore  & istore\_0 & istore\_1 & istore\_2 \\
%		
%		istore\_3 & iload & iload\_0  & iload\_1 \\
%		
%		iload\_2 & istore\_3 & iconst\_m1 & iconst\_4 \\
%		
%		iconst\_5  & iconst\_0 & iconst\_1 & iconst\_2 \\
%		
%		iconst\_3 & bipush  & ifeq  & ifne \\
%		
%		iflt & ifle & ifgt & ifge \\  
%		
%		if\_icmpge & if\_icmple & if\_icmpeq & if\_icmplt \\
%		
%		if\_icmpgt & iinc &	iadd  & isub \\
%		
%		imul & idiv & irem & ineg \\
%		\hline	
%	\end{tabular}
%	
%\end{table}

\figref{fig:holeTimeFD} and~\figref{fig:holeTimeJB} show {\tool} scales with the sketches with increasingly more holes. We do this study for forward differentiation (b10) and bytecode interpreter (b11). As can be seen, {\tool} scales very well. On the other hand, \textsc{AllAtOnce} works well for small instances but soon blows up, timing out on all further instances. The interesting jump in b10 (at \#Holes=8) was seen when we started adding holes for the definitions of the more complex operators like \texttt{sin()} and \texttt{cos()}.

%\subsection{Scaling with counterexamples} \todo{need to add a paragraph as \#cex shown in table~\ref{tab:timing}}
% Line plot of all benchmarks (all holes) in same graph with different colors, x-axis time, y-axis \# cexs

%Separate graphs for (1 hole, 2 holes, all holes) 

\subsubsection{Scaling with size of grammar} 
%\vspace*{-1em}
%Line plot of all benchmarks in same graph with different colors, x-axis time, y-axis \# grammar size
%Separate graphs for (1 hole, 2 holes, all holes) 
Table~\ref{tab:desc} shows that \textsc{IncrementalSynthesis} scales well with the size of the grammar (by the number of productions). On the other hand, \textsc{AllAtOnce} works well  for the benchmarks b1 and b2 as number of holes is only one, but times out for the rest of the benchmarks.
% \textsc{AllAtOnce} is even better than \textsc{Incremental} for small grammars as the latter has to iterate multiple times, firing multiple (albeit simpler) synthesis queries while the former synthesizes definitions for all holes with a single call to the synthesizer.
 
 %We can not relate between time taken to solve the holes with the number of production. What we can conclude that with more increase in the production, there is a possibility of having more holes.\\

The complexity of \textsc{IncrementalSynthesis} is independent of the size of the attribute-grammar but dependent on the length of derivations and the size of the semantic actions. The current state of synthesis-technology allows {\tool} to synthesize practical attribute grammars that have a large number of productions but mostly ``small" semantic actions and where short derivations can ``cover" all productions. Further, any improvement in  program-synthesis technology automatically improves the scalability of {\tool}.

\section{Case Study}
We undertook a case-study on the parser specification of the \textsc{Spin}~\cite{spinRepo} model-checker. \textsc{Spin} is an industrial-strength model-checker that verifies models written in the \textsc{Promela}~\cite{promelaLink} language against linear temporal logic (LTL) specifications. \textsc{Spin} uses \textsc{Yacc}~\cite{yacc} to builds its parser for \textsc{Promela}. The modelling language, \textsc{Promela}, is quite rich, supporting variable assignments, branches, loops, arrays, structures, procedures etc. The attribute grammar specification in the \textsc{Yacc} language is more than 1000 lines of code (ignoring newlines) having 280 production rules.

The semantic actions within the attribute grammar in the \textsc{yacc} description handle multiple responsibilities. We selected two of its operations:

\paragraph{Constant folding array indices} As the \textsc{Promela} code is parsed, semantic actions automatically constant-fold array indices (see Fig.~\ref{fig:constFold}). We removed all the actions corresponding to constant-folding by inserting 8 holes in the relevant production rules (these correspond to the non-terminal \texttt{const\_expr}). The examples to drive the synthesis consisted of \textsc{Promela} code with arrays with complex expressions and the target output was the optimized \textsc{Promela} code. {\tool} was able to automatically synthesize this constant-folding optimization within less than 4 seconds.

\paragraph{AST construction} A primary responsibility of the semantic analysis phase is to construct the abstract syntax tree~(AST) of the source \textsc{Promela} code. We, next, attempted to enquire if {\tool} is capable of this complex task.

In this case, each example includes a \textsc{Promela} code as input and a tree (i.e. the AST) as the output value. We removed the existing actions via 23 holes. These holes had to synthesize the end-to-end functionality for a production rule with respect to building the AST: that, the synthesized code would decides the type of AST node to be created and the correct order of inserting the children sub-trees.

%\begin{wrapfigure}{r}{.4\linewidth}
%	\vspace*{-1cm}
%	\begin{verbatim}
%	struct node{
%	  int val;
%	  char type;
%	  node left;
%	  node right;
%	  ...
%	}
%	\end{verbatim}
%	\caption{structure of an AST node \label{fig:astNodePromela}}
%	%\vspace*{-2cm}
%\end{wrapfigure}

Run of the example suite on the sketchy productions generates a set of programs (one such program shown in Fig.~\ref{fig:promelaAst}); these programs produce \textit{symbolic} ASTs that non-deterministically assigns type to nodes and assigns the children nodes. We leverage the support of references in Sketch to define self-referential nodes.%; for example, a \texttt{node} with \texttt{left} and a \texttt{right} children can we written as shown in Fig~\ref{fig:astNodePromela}.

\begin{figure}
	\begin{subfigure}[b]{.45\linewidth}
		\includegraphics[scale=0.38]{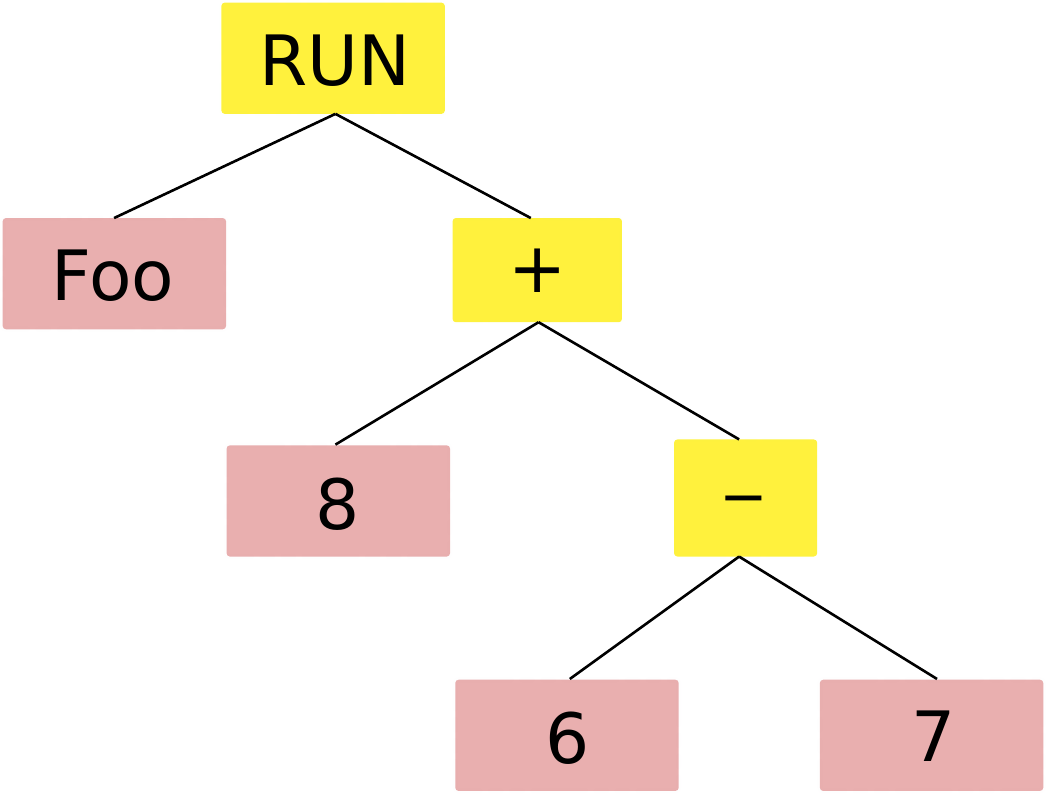}
		\caption{Desired concrete AST \label{fig:codeAst}}
	\end{subfigure}
	\begin{subfigure}[b]{.45\linewidth}
		\includegraphics[scale=0.38]{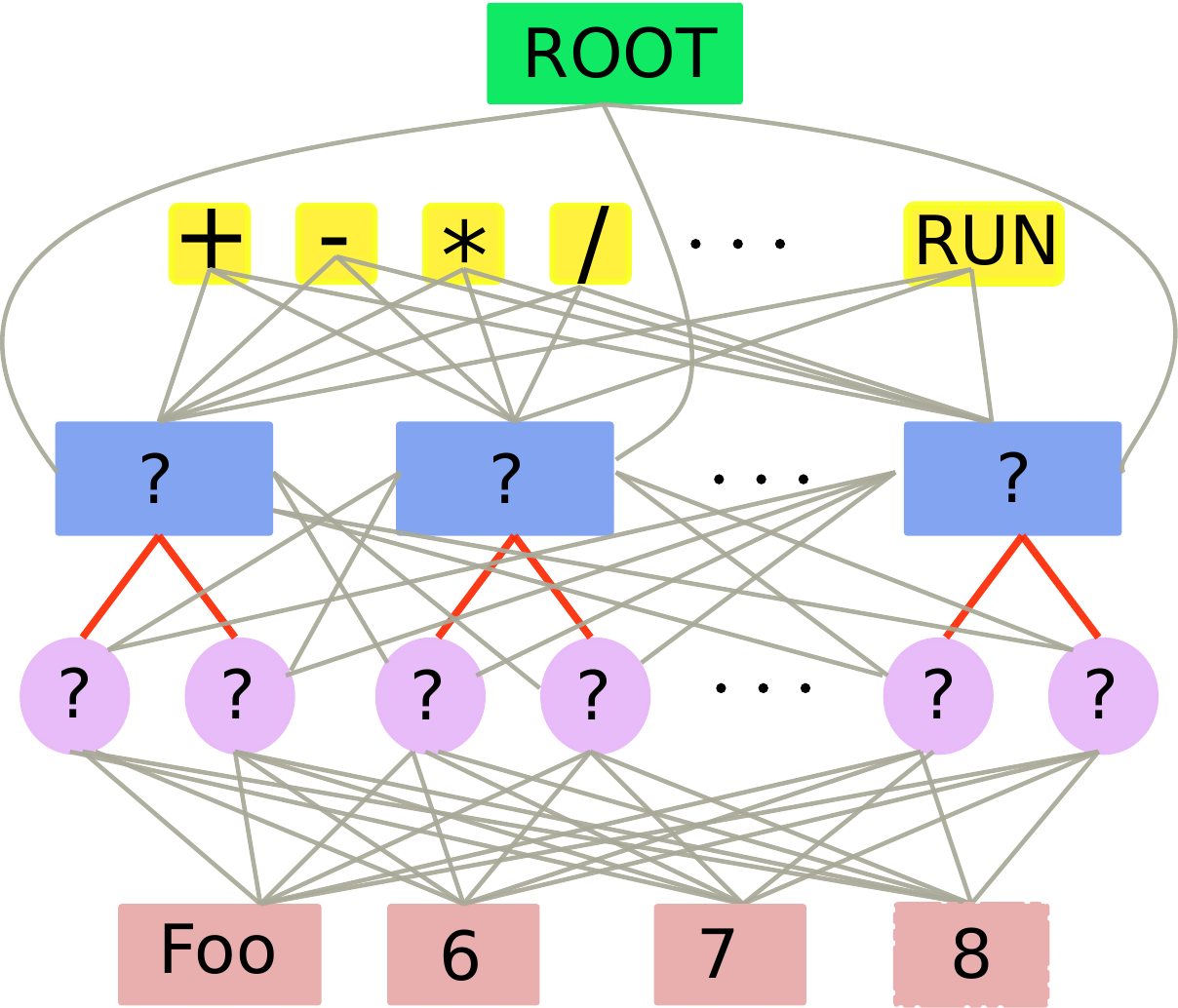}
		\caption{\label{fig:symbolicAst}Symbolic AST}
	\end{subfigure}
				\caption{Desired AST for code in Fig.~\ref{fig:promelaAst} and symbolic AST. Grey lines (in Fig.~\ref{fig:symbolicAst}) denote symbolic choices. \label{fig:astFigs}}
\end{figure}

We insert constraints that establishes tree isomorphism by recursively matching the symbolic ASTs with the respective output ASTs (available in example suite); for example, in Fig.~\ref{fig:astFigs} isomorphism constraints are enforced on the concrete and the symbolic ASTs. Sketch resolves the non-determinism en route to synthesizing the relevant semantic actions. In this case-study, {\tool} was able to synthesize the actions corresponding to the 23 holes within 20 seconds.

%\begin{figure}
%\begin{subfigure}[b]{.4\linewidth}
%%	\hspace{-1cm}
%\begin{lstlisting}[basicstyle=\ttfamily]
%init {
% run Foo(8+(6-7));
%}
%\end{lstlisting}
%		\caption{\textsc{Promela} source code with function call\label{fig:promelaAst1}}
%	\end{subfigure}
%%	\hspace{-2cm}
%	\begin{subfigure}[b]{.4\linewidth}
%		\begin{lstlisting}[basicstyle=\ttfamily,escapeinside={(*}{*)}]
%node n1 = node(val=8);
%node n2 = node(val=6);
%node n3 = node(val=7);
%node n4 = (*\colorbox{yellow}{$\hole{h_a}$}*)(n2, n3);
%node n5 = (*\colorbox{yellow}{$\hole{h_b}$}*)(n1, n4);
%node n6 = (*\colorbox{yellow}{$\hole{h_c}$}*)(`Foo', n5);
%		\end{lstlisting}
%		\caption{Trace generated for code in~\ref{fig:promelaAst1}  \label{fig:promelaAst2}}
%	\end{subfigure}
%	\caption{Trace generation for \texttt{Promela} code to create AST. \texttt{minus, plus} and \texttt{run} are holes in the trace which {\tool} will synthesize.\label{fig:promelaAst}}
%\end{figure}

\section{Related Work}
\vspace*{-.5em}
%\section{Conclusion}\todo{TODO}
Program synthesis is a rich area with proposals in varying domains: bitvectors~\cite{loopFree,PLDI:Solar05}, heap manipulations~\cite{synbad13, synlip15,wolverine17, wolverine21, suslik}, bug synthesis~\cite{apocalypse18}, differential privacy~\cite{kolahal,dpgen}, invariant generation~\cite{closedbox22}, Skolem functions~\cite{manthan20, manthan21a, manthan21b},  synthesis of fences and atomic blocks~\cite{gambit} and even in hardware security~\cite{holl22}.
However, to the best of our knowledge, ours is the first work on automatically synthesizing semantics actions for attribute grammars. 

There has some work on automatically synthesizing parsers: \textsc{Parsify}~\cite{parsify} provides an interactive environments to automatically infer grammar rules to parse strings; it is been shown to synthesize grammars for Verilog, Tiger, Apache Logs, and SQL. \textsc{Cyclops}~\cite{cyclops} builds an encoding for \textit{Parse Conditions}, a formalism akin to \textit{Verification Conditions} but for parseable languages. Given a set of positive and negative examples, \textsc{Cyclops}, automatically generates an LL(1) grammar that accepts all positive examples and rejects all negative examples. Though none of them handle attribute grammars, it may be possible to integrate them with {\tool} to synthesize \textit{both} the context-free grammar and the semantic actions. We plan to pursue this direction in the future.

We are not aware of much work on testing attribute grammars. We believe that our \textit{derivation coverage} metric can also be potent for finding bugs in attribute grammars, and can have further applications in dynamic analysis~\cite{kiteration09,pertinent13,streams13} and statistical testing~\cite{ulysis20, Modi13} of grammars. However, the effectiveness of this metric for bug-hunting needs to be evaluated and seems to be a good direction for the future.

%There has also been work on  equivalence testing of context-free grammars~\cite{madhavanOOPSLA15}. However, we are not aware of much work on testing attribute grammars. We believe that our \textit{derivation coverage} metric can also be potent for finding bugs in attribute grammars. However, the effectiveness of this metric for bug-hunting needs to be evaluated and seems to be a good direction for the future.

\balance
\bibliographystyle{IEEEtran}
\bibliography{refs}

%\clearpage

%\input{chapters/appendix}
\end{document}